\documentclass[journal]{IEEEtran}
\usepackage[pdftex]{graphicx}
\usepackage{amsmath}
\usepackage{color}
\usepackage{soul}
\usepackage{ amssymb }
\begin{document}

\title{A Comparative Taxonomy and Survey of Public Cloud Infrastructure Vendors}

\author{Dimitrios~Sikeridis, Ioannis~Papapanagiotou,   Bhaskar~Prasad~Rimal, and~Michael~Devetsikiotis
\thanks{Dimitrios Sikeridis, Ioannis Papapanagiotou, Bhaskar Prasad Rimal, and Michael Devetsikiotis are with the Department of Electrical and Computer Engineering, The University of New Mexico, Albuquerque,
NM, 87131 USA, e-mail: \{dsike, ipapapa, bhaskar, mdevets\}@unm.edu}
}

\maketitle

\begin{abstract}

An increasing number of technology enterprises are adopting cloud-native architectures to offer their web-based products, by moving away from privately-owned data-centers and relying exclusively on cloud service providers.
As a result, cloud vendors have lately increased, along with the estimated annual revenue they share. However, in the process of selecting a provider's cloud service over the competition, we observe a lack of universal common ground in terms of terminology, functionality of services and billing models. This is an important gap especially under the new reality of the industry where each cloud provider has moved towards his own service taxonomy, while the number of specialized services has grown exponentially. This work discusses cloud services offered by four dominant, in terms of their current market share, cloud vendors. We provide a taxonomy of their services and sub-services that designates major service families namely computing, storage, databases, analytics, data pipelines, machine learning, and networking. The aim of such clustering is to indicate similarities, common design approaches and functional differences of the offered services. The outcomes are essential both for individual researchers, and bigger enterprises in their attempt to identify the set of cloud services that will utterly meet their needs without compromises. While we acknowledge the fact that this is a dynamic industry, where new services arise constantly, and old ones experience important updates,  this study paints a solid image of the current offerings and gives prominence to the directions that cloud service providers are following.

\end{abstract}


\IEEEpeerreviewmaketitle

\section{Introduction}

Cloud computing has introduced a modern computing and storage paradigm, by virtualizing the hardware along with the software, and providing it as a service over the Internet \cite{armbrust2010view, vaquero2008break}. This new paradigm does not burden the customer (either a company or an individual) with server interaction and relies on cloud service providers for the maintenance and management of the resources. In exchange, customers are charged under a pay-per-usage billing model that has lead to a whole new class of technology providers. During the last five years, the advances in cloud computing have managed to defy all predictions both in terms of advancing various related technologies (databases, networking, machine learning applications) but also by creating new markets and profits for the cloud-related companies, and vendors \cite{bartels2014public}. 

In light of this, relying on their enormous existing computing infrastructure used to host their own services, major technology companies such as Amazon, Microsoft, and Google started renting the capacity of their data centers to companies and individuals around the world \cite{prodan2009survey}. By doing that, they provided efficient and scalable computing centers that other companies could not develop or maintain on their own. This combination of low-cost (zero installation and maintenance costs) and high-performance (high-end supercomputers with unlimited memory capabilities) was quickly accompanied with other advantages,  including minimal cost of software, unlimited storage capacity, data reliability, universal data access and multiple user collaboration. Therefore, the transition to cloud based services became a viable solution for the majority of technology companies and a must for any new start-up. Interestingly, more and more enterprises abandon nowadays the paradigm of huge private-owned data-centers and move the entirety of their services to the cloud transforming to totally cloud-native companies.  

This long list of advantages offered by cloud computing has lead to a growing number of competitors in the cloud industry.
Company executives interested in investing to lead their business to the cloud era should have a clear understanding of each vendor's offerings and how they can utilize each cloud environment to successfully serve their needs. There is a number of research works that focus on the selection of cloud service providers mainly by using the service level agreements (SLAs) \cite{faniyi2016systematic}, that guarantee the provided quality of services or customer satisfaction-based measures \cite{qu2013cloud, sun2014cloud, esposito2016smart, ghosh2015selcsp}. However, today the cloud services offered have been actively reformed in comparison with previous years \cite{prodan2009survey, li2010cloudcmp}. Consequently, the interested companies or individuals need to focus on comparing the specific offerings of all cloud service providers in order to make the optimal for them decision in terms of overall cost, and offered service types that fulfill their needs without compromises.

In this work, we perform a taxonomy of the services provided by the cloud players. Moreover, we briefly compare mature key services that are regularly utilized by cloud applications. The focus is on identifying common architectures, functionality, terminology, and possible open research issues. Since there are many different companies offering similar services, we will focus on the top four, sorting them by their current market share. We identify several major groups and categories of offered services:
\begin{itemize}
\item{Compute Services}
\item{Storage Services }
\item{Databases}
\item{Big Data and Analytics}
\item{Data Pipelines}
\item{Machine Learning and Artificial Intelligence}
\item{Networking and Content Delivery}
\end{itemize}
For each category, we will discuss and compare only the services offered by the four dominant vendors. However, there will be references to additional and innovative services that some vendors exclusively offer ahead of the competition. We will mainly focus on the functionality and features offered by each service making minor commends about the pricing as it can get quite convoluted. However, it is an important issue to examine separately, as it can be the decisive factor to consider before favoring a vendor over the competition \cite{spectator},\cite{megaguide}. 

Moreover, this work captures a specific time frame of the industry services' state and utilizes technical information of the surveyed providers made available until the end of 2017.  We try to focus on describing mature services that have been established as standard products during a relatively large period. Since this is a competitive market in the heart of technology innovations, all services are updated continuously, and new ones are rapidly released every year. However, we believe that this taxonomy and survey constructs a solid image of the modern cloud industry, and highlights its trends. To our best knowledge, this classification that reflects the current conditions and more importantly the current customer needs, has not been recently attempted mainly due to the dynamic nature of the cloud computing market.

The remainder of the paper is structured as follows. Modern cloud deployment scenarios and dominant service providers are briefly described in Section~\ref{section:vendors}. Section~\ref{section:compute} presents the computing services. Storage services and cloud databases are discussed in Sections~\ref{section:storage}, and~\ref{section:database} respectively. Sections~\ref{section:ai}, and~\ref{section:pipe}  present services related to big data, analytics and data pipelines. Section~\ref{section:ma} discusses services that support machine learning and artificial intelligence applications. Section~\ref{section:net} discusses network-related services, while section~\ref{section:rest} briefly presents additional cloud services. Finally, Section~\ref{section:chal} briefly highlights related research challenges, and  Section~\ref{section:con} concludes the paper.

\section{Cloud Deployment Scenarios and Vendors}
\label{section:vendors}

\begin{figure}[t]
\centering
\includegraphics[width=0.85\columnwidth]{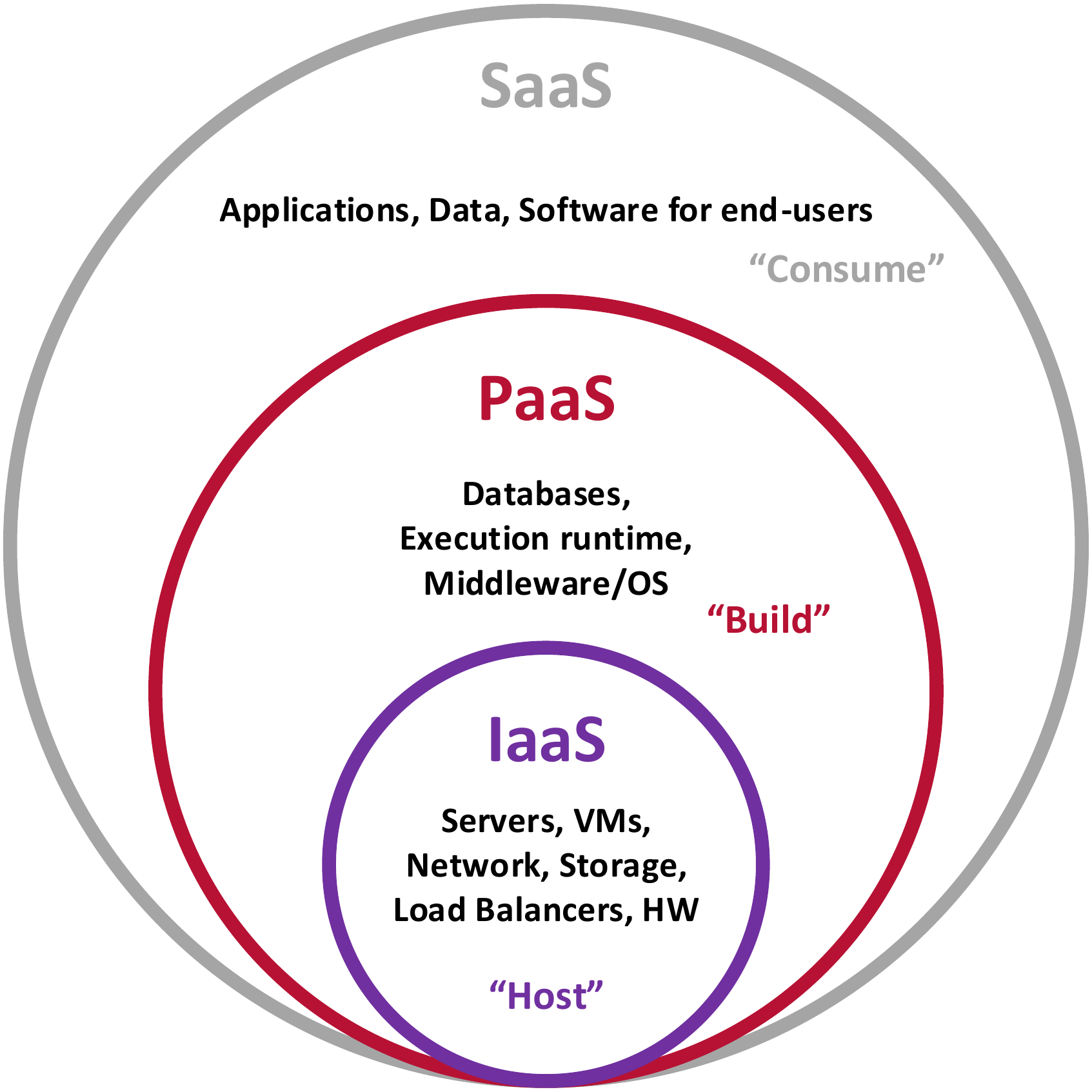}
\caption{Cloud Computing Solutions} \label{fig:aas}
\vspace{-0.4cm }
\end{figure}

Cloud vendors offer typically three public cloud layers and deployment scenarios \cite{armbrust2010view, rimal2009taxonomy}, as depicted in Fig. \ref{fig:aas}:
\begin{itemize}
\item {\it Infrastructure-as-a-Service (IaaS):} enables the client to build and manage databases and applications using the virtual servers, storage space, and hardware of the vendor's data center. The core in this scenario is hardware virtualization that allows deployment of guest operating systems and applications on top of remote equipment resulting to scalable, distributed solutions. Moreover it provides on-demand services to clients using a
shared platform architecture and offering increased flexibility.

\item {\it Platform-as-a-Service (PaaS):} enables the client to build and manage applications while the vendor hosts the hardware and software on its own infrastructure. In addition to the hardware included in IaaS deployments, PaaS includes development tools, management systems, middleware and any other tool required for building, testing, and fully distributing a web application.

\item {\it Software-as-a-Service (SaaS):} is typically built on top of a PaaS cloud solution, whether that platform is publicly available or not, and provides software for end-users. It is a relatively restrictive model, where customers utilize pre-designed services instead of deploying their own.

\end{itemize}

\subsection{Cloud Vendors and Industry Market Share}

During the period 2012-2015, cloud computing was responsible for 70\% of the related IT market growth. The total amount of revenues for the related public and privet cloud services (hardware, software, middleware) reaches annually the 50 billion dollars mark \cite{revenue}. The market is still growing with Cisco predicting that by 2020, 92\% of related workloads will be
processed by cloud data centers, while only 8\% will be processed by traditional data centers. The same report \cite{index2016forecast} also analyses the predictions of installed workload which by 2020 will be massively leaning towards SaaS workloads. Finally, it is predicted that in three years hyper-scale data centers will grow double in numbers as they will represent 47\% of all data center servers.

 \begin{table*}[t]
\scriptsize
\renewcommand{\arraystretch}{1.3}
\caption{Compute Services}
\label{Table:compute}
\centering
\begin{tabular}{||l|c|c|c|c||}
\hline 
\bf{Service Type}   & \bf{Amazon}&  \bf{Microsoft}& \bf{Google}& \bf{IBM}    \\ \hline \hline
Virtual Machines &AWS EC2 & Azure Virtual Machines & Compute Engine  & Virtual Server \\ 
          &         & Azure Virtual Machine Scale Sets  &                 & Bare Metal Servers\\ \hline
Auto-scaling & AWS Auto Scaling   & Azure Autoscale  & Compute Engine Autoscaler  & Bluemix Auto-scaling \\ \hline
Container Image Registry & AWS EC2 Container Registry   & Azure Container Registry   & Container Registry  &  Container Registry Archives\\ \hline
Container Service & AWS  EC2 Container Service   & Azure Container Service    & Container Engine  &  Container Service \\ \hline
Serverless Computing & AWS Lambda   & Azure Functions  & Cloud Functions  & OpenWhisk \\ \hline
\end{tabular}
\end{table*}
 
A vast number of companies compete to establish themselves as leaders and innovators of the cloud industry. Without a doubt, Amazon's AWS \cite{aws} is leading the global public cloud market after ten years of its launch. Today AWS platform is utilized by over 1 million organizations of various sizes as well as independent cloud developers and government entities providing application, and infrastructure services, including VMs, storage and content delivery, networking, API management and a number of management, security, or migration tools. AWS is retaining its dominant share of the growing public cloud services market being over 40\%, while recent evaluations of global public cloud platform providers - Forrester \cite{forrester}, Synergy \cite{synergy} - agree that the next three chasing providers, -Microsoft with Azure, Google with Google Cloud Platform and IBM with Bluemix- are steadily gaining ground at the expense of smaller providers in the market. In aggregate, these three vendors account for 23\% of the total public IaaS and PaaS cloud market, at the moment, and seriously challenge the dominance of AWS in the field. 

Microsoft with Windows Azure \cite{azure} has made considerable steps to gain a leadership position in the market, second only to AWS. Apart from own products, Microsoft has managed to offer a variety of external open source tools, services, and platforms on top of their big collection of infrastructure and application services \cite{forrester}. Google \cite{google} is a very interesting case, regarding its involvement in cloud computing. Apart from owning the platform in terms of data centers and infrastructure between data centers, they are deeply involved in developing "edge" devices such as Google smartphones running native Android software, or Google Home devices equipped with their own APIs. This involvement enables Google to access, control and transfer massive data very fast. This is a significant advantage and combined with their renewed focus on the enterprise (by offering stronger security and machine learning services with a reduced pricing strategy) they can escalate their growth.  Finally, IBM Cloud includes their Bluemix developer platform \cite{blue}, which is the base of their cloud services, and the SoftLayer IaaS service \cite{soft}. Their points of strength include application migration, cognitive analytics (e.g., Watson Platform) and the ability to host complex hybrid cloud formations targeting enterprises in cloud transition. Their ongoing challenge is to combine SoftLayer and Bluemix services into a single platform in order to offer a consistent hosting platform.

\subsection{Other Major Cloud Providers}
For the needs of this work, we will compare the services of the four aforementioned market leaders. However, smaller market players should not be underestimated or written off. On account of completeness, we will make a brief reference to the rest quickly growing cloud vendors.

{\it Oracle} may lack the functionality scale of the others but has made a huge leap by providing a dependable cloud platform using its strong development experience. They target existing clients and with their strong commitment to public cloud platforms, they continue to grow aiming for a global presence in the next two years \cite{forrester}.

{\it VMware} is a vendor that has recently attempted to take their computing environment and optimize it to run on bare metal AWS infrastructure. That way they have introduced a new market model for customers with a promising future.

{\it Rackspace} is implementing an actual commercialization of the OpenStack cloud operation system \cite{openstack}. OpenStack has been very successful being supported by a large number of IT enterprises (e.g., RedHat, Cisco, HP) giving a chance to smaller vendors to compete with Amazon. Essentially, OpenStack is a large set of open-sourced software tools designed to use pooled virtual resources for building and managing public and private cloud platforms. The associated tools implement the core of cloud services (compute, storage, networking, identity, image). In addition, optional functions that can easily be developed to meet any needs due to the open-source nature of the architecture.

{\it Joyent} is a vendor offering cloud infrastructure and analytics services utilizing an architecture that fundamentally changes the economics of cloud computing (public and private). Their approach utilizes containers (see Section \ref{section:compute}-B) running directly on top of bare-metal servers avoiding the complexity of managing virtual machines in the traditional sense, and providing increased performance with reduced cost. In addition, the company established the use of Node.js to the industry offering exclusive debugging and performance tools for Node.js applications. In 2016 Joyent was acquired by Samsung.

Finally, {\it Salesforce} is known for their SaaS cloud services being a platform of public cloud that primarily configures, extends, and integrates SaaS products (e.g., Customer Relationship Management (CRM)), while {\it  CenturyLink} that entered the market in 2013, offers mostly infrastructure services — multitenant IaaS, bare metal, and dedicated private cloud — from 60 global data centers and primarily targets clients shifting their infrastructure to the cloud \cite{forrester}.

\begin{table*}[!ht]
\scriptsize
\renewcommand{\arraystretch}{1.3}
\caption{Virtual Machines}
\label{Table:vs}
\centering
\begin{tabular}{||l|c|c|c|c||}
\hline 
\bf{Virtual Machine Services \& Features}   & \bf{Amazon}&  \bf{Microsoft}& \bf{Google}& \bf{IBM}    \\ \hline \hline
Service & AWS EC2 & Azure Virtual Machines & Compute Engine  & Virtual Server \\ 
               &         & Azure Virtual Machine Scale Sets  &                 & Bare Metal Servers\\ \hline
Hypervisor     & Xen     & Hyper-V               & Hyper-V         &  Xen \\  \hline
Auto Scale Ability     & \checkmark      & \checkmark                & \checkmark      &  \checkmark  \\  \hline
Max vCPUs      & 128 (X1)     & 32 (G5)          & 64        &  56 \\  \hline
Max Memory (GB) & 1952 (X1)     & 448 (G5)        & 416 (6.5/vCPU)   &  242 \\  \hline
Max Attached Storage (GB) & 3840 (X1)     & 6598 (G5)        & 4096 (64/vCPU)   &  100 \\  \hline
Max Instance Storage (GB) & 48000 (D2)     & 32000         & 64000    &  - \\  \hline
Custom VMs               &                &                   & \checkmark  &  \checkmark \\  \hline
Dedicated Hosts  & \checkmark     &         &   & \checkmark \\  \hline
Bare Metal       &   \checkmark       &         &   &  \checkmark \\  \hline
\end{tabular}
\end{table*}

\section{Compute Services}
\label{section:compute}

This family of services provides the core of processing and calculating capabilities along with the power to run applications in the cloud environment. Over the years enterprise computing has slowly shifted from virtual machine-based architectures to the utilization of containers and nowadays towards serverless computing \cite{baldini2017serverless}.
This shifting was initiated due to the size and complexity of VMs that follows their need to include all the components required by multiple applications. On the contrary, container-based systems (such as Docker) depend on virtual memory hardware without the virtual machine support to host applications. A single container instance can either run on a VM or completely alone on top of an operating system. Containers offer a more flexible security environment and abstract the development of the underlying operating system. They usually depend on less code and have less computing overhead compared to applications running in a VM. In 2014 Amazon introduces AWS Lambda which is essentially the first serverless cloud-compute service. The serverless architecture treats a single application as a set of different functionalities (or services) which are usually triggered by events. The developers just provide their function code, and attach a related event source, with the cloud provider taking care of provisioning, deploying, and managing all the sufficient computing resources to support the application code. This new paradigm often referred to as Function as a Service (FaaS) introduces many advantages including reduced application hosting costs (pay-per-execution policies, zero costs for idle time, low maintenance and administration costs), complete abstraction from hardware (developers can purely focus on their application's code and functionality), and better support for the emerging category of event-driven applications. Table \ref{Table:compute} summarizes the services offered by each vendor in this vast computing services category.

\subsection{Virtual Machines}

Amazon's basic compute component is AWS Elastic Compute Cloud (EC2) that provides over 40 different sizes of instances/virtual machines. These instance types provide different optimization combinations concerning CPU, storage, memory or networking performance to provide flexibility for any application need. Some of these instances can address computationally and memory intense applications (e.g., X1 instances), while others are designed for cases where high capacity is preferred offering up to 48 TB of additional attached storage (D2 instances). Finally, Amazon offers EC2 Bare Metal Instances enabling applications to utilize directly physical hardware resources of the AWS infrastructure.

Microsoft Azure has a similar approach concerning their instances and the associated service is called Azure Virtual Machines (VMs). A wide variety of different use cases is covered with the client being able to deploy up to 32 vCPUs with up to 448 GB of memory and attached maximum storage of 6144 Gibibytes (=6598 Gigabytes, GiB=$1024^{3}$ bytes) (G5 VM type). A recent compute service addition from Microsoft is Azure Virtual Machine Scale Sets, where the user is able to set up and manage multiple identical Virtual Machines. The service allows the deployment of up to 1000 VMs for Microsoft-provided configurations and up to 100 for custom purposes. Possible benefits include increased availability, better cost management, and improved fault tolerance.

The same happens on Google's side with the Google Cloud Service Engine which delivers virtual machines with a variety of types that have a fixed collection of resources. Apart from that, Google gives the opportunity to deploy fully customizable VMs with up to 64 vCPUs, 6.5 GB/vCPU memory size and 64 GB/vCPU maximum attached storage. Moreover, Google Cloud Engine was the first to offer an embedded live migration service that migrates running instances to another host instead of rebooting them when a system event occurs (SW/HW update). However, other vendors have been catching up, with a case in point being Amazon's recently launched AWS Server Migration Service.

IBM's SoftLayer has a different view regarding the virtual servers by offering only custom VMs. The customer is able to configure his own machines using up to 56 vCPUs and up to 242 GB  of memory. Another service of SoftLayer is the Bare Metal Servers, again with a variety of different combinations and offering the advantage of single-tenant configuration generally used for security purposes (similar to AWS's EC2 Dedicated Instances). Table \ref{Table:vs} summarizes the key features of the four vendor's virtual machine services.

One important functionality concerning - and practically defining - cloud computing is the automatic addition or removal of instances/virtual machines from a managed instance group based on the fluctuation of the load (increase/decrease). Scaling the infrastructure to meet the changing demand is really important, mainly for saving capital by optimizing performance at the same time. All four aforementioned vendors offer seamless and automatic scaling to follow the demand: AWS with AutoScaling Service, Microsoft with Azure Autoscale, IBM with SoftLayer/Bluemix Auto-Scaling and finally Google with Compute Engine Autoscaler and by fully integrating autoscaling into its monitoring solution Google Slackdriver.

\begin{table*}[ht]
\scriptsize
\renewcommand{\arraystretch}{1.3}
\caption{Serverless Computing}
\label{Table:event}
\centering
\begin{tabular}{||l|c|c|c|c||}
\hline 
\bf{Serverless Computing Services \& Features}   & \bf{Amazon}&  \bf{Microsoft}& \bf{Google}& \bf{IBM}    \\ \hline \hline
Service & AWS Lambda   & Azure Functions  & Cloud Functions  & OpenWhisk \\ \hline
Supported Languages     & Python, Java,     & Python, Javascript,   & Javascript  &  Python, Javascript,  \\ 
                        &        Javascript            &  C\#,  PHP        &             & (Swift, Docker)  \\\hline
Max Execution Time / Request   & 5 min     & Unlimited          & Unlimited        & 5 min   \\  \hline
Scalability     & Automatic scaling    & Automatic scaling    & Automatic scaling   &  Automatic scaling \\  \hline
HTTP Invocation & API Gateway     & HTTP Trigger       & HTTP Trigger   &  API Gateway   \\  \hline
Log Management & Cloud Watch  &  Kudu Console    & Stackdriver Logging   &  Bluemix UI / Cloud Foundry CLI  \\  \hline
Concurrent Executions &100 parallel  &10 instances        & Not Specified  &  Not Specified \\  \hline
\end{tabular}
\end{table*}

\subsection{Container Services}

\begin{figure}[t]
\centering
\includegraphics[width=\columnwidth]{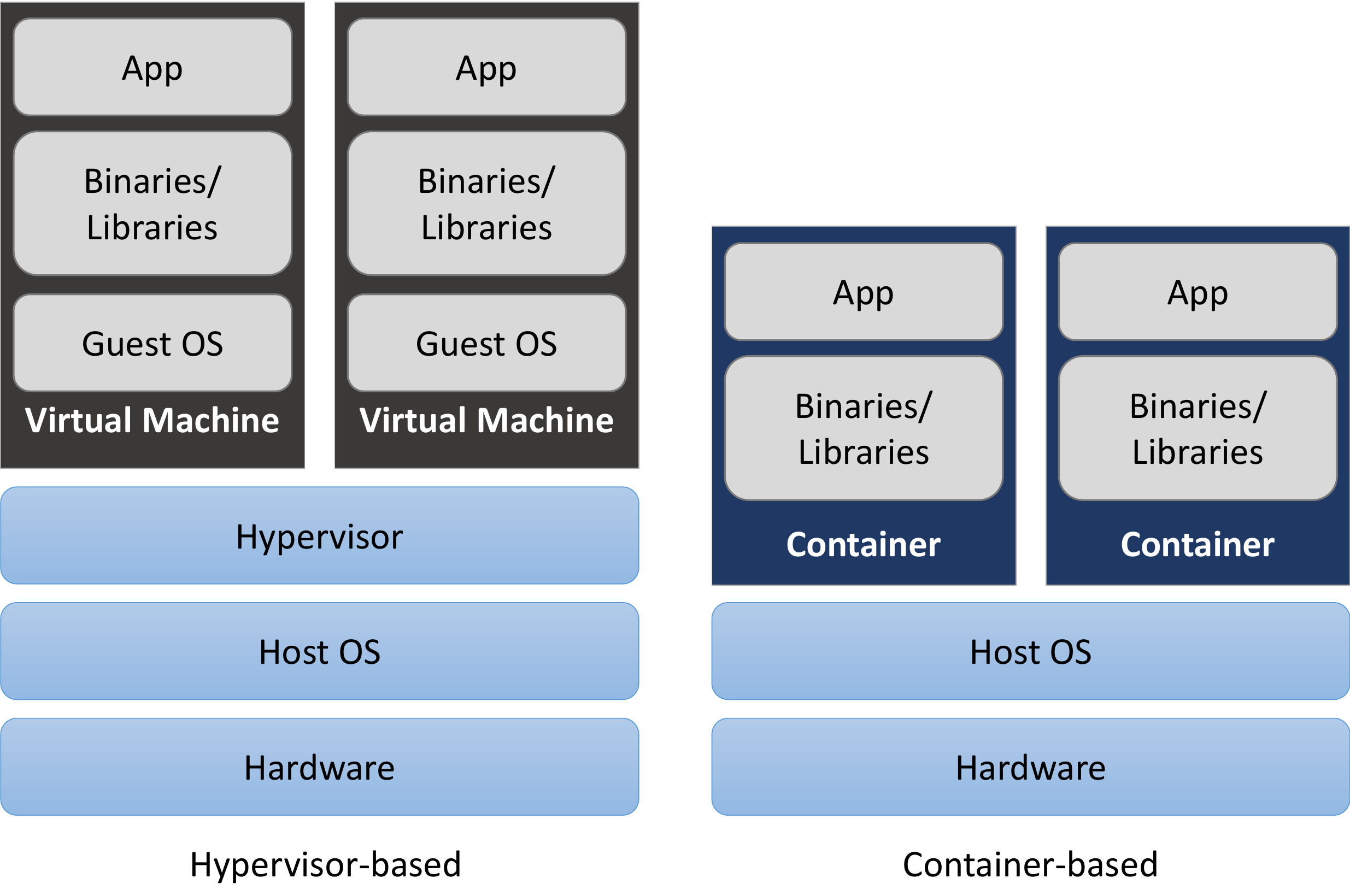}
\caption{Hypervisor-based vs Container-based Virtualization } \label{fig:cont}
\vspace{-0.4cm }
\end{figure}

Container-based visualization follows a different approach than hypervisor-based visualization by removing an operating system - an additional software - layer and sharing the host system's kernel as shown in Fig. \ref{fig:cont} \cite{pahl2017cloud, kozhirbayev2017performance}. It is becoming a frequently-used virtualization solution for PaaS and IaaS clouds due to containers' increased density, isolation, elasticity, and rapid provisioning. Containerization uses lightweight packages instead of full VMs, and simplifies migration of applications between private, public, and hybrid clouds.

Containers - despite being an old technology - are increasingly used to boost cloud application portability and efficiency with Docker \cite{fink2014docker} being a leader in the field. Docker is an open-source project that utilizes software containers to automate the deployment
of applications, and at the same time provides an additional layer of
abstraction of operating system-level virtualization on Linux.

Regarding container management and control, Container Orchestration Environments (COE) offer features such as provisioning, scaling, managing dependencies, handling updates or failures, enabling discovery and interoperability. There is a number of open-source COE tools designed to manage containers when running multiple instances of an application. Docker's COE offering is Docker Swarm while other alternatives include 
Kubernetes \cite{burns2016borg, brewer2015kubernetes} - an open-source container manager originally developed by Google- and Mesosphere's DC/OS Marathon. The latter offering is built on top of Apache Mesos \cite{hindman2011mesos}, which is a manager of clusters designed for efficient resource isolation and sharing across distributed frameworks, applications, or hosts.

Following the aforementioned opportunities and functionality provided by containers, it is not a surprise that all four vendors have \textit{Container-as-a-Service (CaaS)} offerings. Amazon's solution is AWS EC2 Container Service (ECS), a full container management service that supports Docker containers. These containers are exclusively run on AWS EC2 instances and the created clusters are coordinated through the Amazon ECS Container Agent executed on each EC2 instance inside the cluster. Container execution on infrastructure outside EC2 is not supported but ECS's strength lies with the strong integration with the rest of AWS services (such as AWS Identity and Access Management (IAM), AWS CloudTrail to provide metrics and container logging, AWS CloudFormation or AWS Elastic Load Balancers). Moreover, apart from the custom ECS scheduler and manager, Amazon offers to the users the ability to compose their own schedulers or integrate third party products (like Marathon) through the ECS APIs. Finally, they offer an AWS Container Registry Service (ECR) that enables developers to publish their own docker container images. In addition, the service offers an interface along with APIs for management and connection with the other AWS services. ECR also encrypts docker images (AWS Simply Storage Service (S3) encryption), stores them in S3 for availability and transfers them over HTTPs to ensure protection. An interesting addition by Amazon is the AWS Elastic Container Service for Kubernetes (EKS) which is a cluster manager developed by Google. The new addition supports Kubernetes integration with AWS services and relieves customers from managing scaling and availability of Kubernetes clusters across multiple availability zones.

\begin{table*}[th]
\scriptsize
\renewcommand{\arraystretch}{1.3}
\caption{Storage Services}
\label{Table:storage}
\centering
\begin{tabular}{||l|c|c|c|c||}
\hline 
\bf{Service Type}   & \bf{Amazon}&  \bf{Microsoft}& \bf{Google}& \bf{IBM}    \\ \hline \hline
Object Storage & AWS Simple Storage Service (S3) & Azure Blob Storage & Cloud Storage & Cloud Object Storage\\ \hline
Virtual Machine Disk (Block) Storage  & AWS Elastic Block Storage (EBS) & Azure Managed Disks Storage & Cloud Persistent Disk &  Bluemix Block Storage \\  \hline
Long Term Cold Storage   & AWS Glacier      & Azure Blob Storage (Cool) & Cloud Storage (Coldline) & Object Storage (Cold Vault) \\ \hline
File Storage    & AWS Elastic File System (EFS)      & Azure File Storage & Cloud Storage & Bluemix File Storage\\ \hline
\end{tabular}
\end{table*}

Microsoft offers Azure Container Service (ACS) that incorporates multiple container orchestration tools for the developer to use. Choices include Docker Swarm, Mesosphere DC/OS and lately Kubernetes with Azure having APIs available enabling the use of any other similar tool. ACS supports Linux containers with all orchestrators, while Windows container support with Kubernetes is currently in preview. Furthermore, the service integrates other Azure tools like Resource Management with features that define configurations. Apart from that, Azure Container Registry is offered as a managed container image store and management service. Developers can pull container images, push them for storage purposes and use the aforementioned orchestration systems or other Azure services (e.g., Container service, Batch, Service Fabric).

Google Container Engine (GCE) is used to run docker containers and is powered by Kubernetes, Google's own open-source cluster manager. Kubernetes runs a master node outside each project to coordinate the hosts that run on instances inside the project. Moreover, GCE is also integrated with the other services for container metrics and utilizes a JSON-based syntax to define the hosts' behaviour before Kubernetes handles the cluster monitoring. Google, finally, unlike the other providers offers by default a private docker registry with the Google Container Registry service.

The last of the cloud vendor to launch integrated container cluster management was IBM with its own Bluemix Container Service, a docker-based platform that provisions, updates, and monitors user's containers. Currently, Bluemix Container Service has released a beta of Kubernetes for container orchestration to automate deployment, scaling, and monitoring. Moreover, the service provides completely native Kubernetes APIs, build in security scanning with Bluemix Vulnerability Advisor, automatic load balancing, performance metrics and access to other cloud services, including IoT tools, Watson API, and blockchain. Finally, IBM's stand alone container registry service is the recently added Container Registry Archives. This functionality is also supported by the Bluemix Container service, which provides an image registry that handles the developer's container images.

\subsection{Serverless Compute}

Another important service adopted by cloud vendors is serverless computing. This type of services allows the client to upload his own code with the execution being triggered by predefined events without provisioning or managing servers, compute resources or storage. Thus, the process requires limited effort towards deploying a workload. In addition, when small single-step workload are considered serverless services are less complicated in setting up and launching containers or tasks without presenting many security dependencies and resiliency issues.
Table \ref{Table:event} highlights key features of serverless platforms across all vendors.

Amazon was the first to hit the market with such a service when it introduced Lambda in its 2014 reInvent conference. Lambda functions are now a native part of the AWS ecosystem and can be triggered by HTTP endpoints, in-app activity of mobile apps or through the AWS services, such as S3, Dynamo DB, Kinesis SNS or CloudWatch. Some of the capabilities include real-time stream processing, data validation filtering or sorting, serverless backend support (IoT backend, mobile backend) and web applications support. The service is also able to support AWS Step Functions -enables the development and functionality of distributed applications providing a graphical console to visualize and arrange application components-, along with REST interfaces (API Gateways). Currently, AWS Lambda supports Node.js, Python, and Java.  On the other hand, due to the immaturity of the product cases of limited operational visibility have arisen with companies utilizing novel approaches to manage distributed configurations. Also, as code complexity of applications that rely on the serverless paradigm is increasing, difficulties concerning remote debugging (or even complete lack of it) may occur. 

Microsoft released in May 2016 the Azure Functions service, which is the company's evolution of PaaS programming for custom code execution, supporting C\#, JavaScript, Python, and PHP. Highlighting differences, the Lambda service is organizationally independent while Azure Functions are grouped locally in an "application". Also, Azure Functions memory allocation is performed per app service and not per function as happens in AWS Lambda.

In February 2016, Google introduced the Alpha release of Google Cloud Functions. The platform currently supports JavaScript and event triggering by internal Google Cloud Storage events, Google Cloud Pub/Sub or HTTP invocations.

Finally, IBM released its own beta service, the IBM OpenWhisk \cite{openwhisk} in February of 2016. The language support includes Node.js, Python, Swift, and Docker. Swift and Docker are interesting choices as Swift allows iOS developers to build their own back-ends easily, while Docker allows the implementation of actions in any language. Moreover, IBM offers the ability to connect, chain and reuse client functions while at the same time OpenWhisk supports 3rd party integration. However, with the platform being in a beta release there are some missing features including HTTP customization, lack of versioning, and documentation gaps.

%

\section{Storage Services }
\label{section:storage}

Alongside the compute services, persistent data support is a key features of modern cloud computing. The storage services provide a wide variety of different methods for storing and managing data from an application through the cloud. Table \ref{Table:storage} summarizes the persistent data services provided by the Cloud vendors we are examining.

\subsection{Object Storage}

\begin{table*}[th]
\scriptsize
\renewcommand{\arraystretch}{1.3}
\caption{Object Storage}
\label{Table:os}
\centering
\begin{tabular}{||l|c|c|c|c||}
\hline 
\bf{Object Store}   & \bf{Storage cost}&  \bf{Egress cost}& \bf{Availability (\%)}  & \bf{Regions}   \\ 
                    & \bf{(cents/GB/mo)}    &  \bf{(cents/GB)}                           &                         &                           \\ \hline \hline
Amazon Glacier      & 0.40                  & 9.0                    	  & 99.99 & 12 -USA(4), UK(2), China(1), Japan(1), India(1),   \\
					&						&									&		  &    Australia(1), South Korea(1), Germany(1)-  \\  \hline
Amazon S3 Infrequent Access     & 1.25                   & 10.0                   &  99.90 & 15 -USA(4), UK(2), China(1), Japan(1), India(1),     \\
Amazon S3 Reduced Redundancy     & 2.40                   & 9.0                     & 99.99 &    Australia(1), South Korea(1), Germany(1),  \\ 
Amazon S3 Standard     & 2.30                   & 9.0                     & 99.99 &    Brazil(1), Singapore(1), Canada(1)-                \\ \hline \hline								
					
Microsoft Azure       &                &                    &          & 30 -USA(8), UK(3), China(3), India(3),       \\
Geographically Redundant Storage &	     						&				&		  &   Australia(2), South Korea(2),   \\ 
(Cool)   	 &		2.00    				&	9.7					 				&	99.00	  &   Canada(2),  Japan(2), Germany(2),     \\
(Hot)  	  & 3.68                   & 8.7                     & 99.90  &      Netherlands(1), Brazil(1), Singapore(1)-             \\ \hline


Microsoft Azure       &              &                   &   & 30 -USA(8), UK(3), China(3), India(3),     \\
 Locally Redundant Storage &	               							&				&		  &     Australia(2), South Korea(2),  \\ 
(Cool)  	& 1.00                   & 9.7                    & 99.00  &   Canada(2),  Japan(2), Germany(2),       \\ 	
(Hot)  	 & 1.84                   & 8.7                     & 99.90 & Netherlands(1), Brazil(1), Singapore(1)-             \\ \hline	


Microsoft Azure Read-Access       &                                   &  &  & 30 -USA(8), UK(3), China(3), India(3),    \\
Geographically Redundant Storage&	               						&				&		  &  Australia(2), South Korea(2),    \\ 
(Cool) & 2.50                   & 9.7                     & 99.00     &            Canada(2),  Japan(2), Germany(2),                  \\ 
(Hot) & 4.60                  & 8.7                  & 99.90 & Netherlands(1), Brazil(1), Singapore(1)-             \\ \hline	\hline


Google Cloud Coldline Storage       & 0.70                  & 12.0                   &  99.00 & 8 -USA(4), Europe(1), Asia(3)-   \\ \hline 
Google Cloud Multi-Regional Storage       & 2.60                  & 12.0                   &  99.95 & 8 -USA(4), Europe(1), Asia(3)-    \\ \hline 	
Google Cloud Regional Storage		 & 2.00                  & 12.0                   &  99.90 & 8 -USA(4), Europe(1), Asia(3)-    \\ \hline 	
Google Cloud Nearline Storage		 & 1.00                  & 12.0                   &  99.00 & 8 -USA(4), Europe(1), Asia(3)-   \\ \hline \hline

IBM Bluemix/SoftLayer Object Storage	 & 3.00                 & 9.0                   &  - & 25 -North America(8), Europe(8), Asia \& Pacific(6),     \\ 
&		&	&	  &  Australia(2), South America(1)-                \\ \hline
\end{tabular}
\end{table*}

\begin{figure}[t]
\centering
\includegraphics[width=\columnwidth]{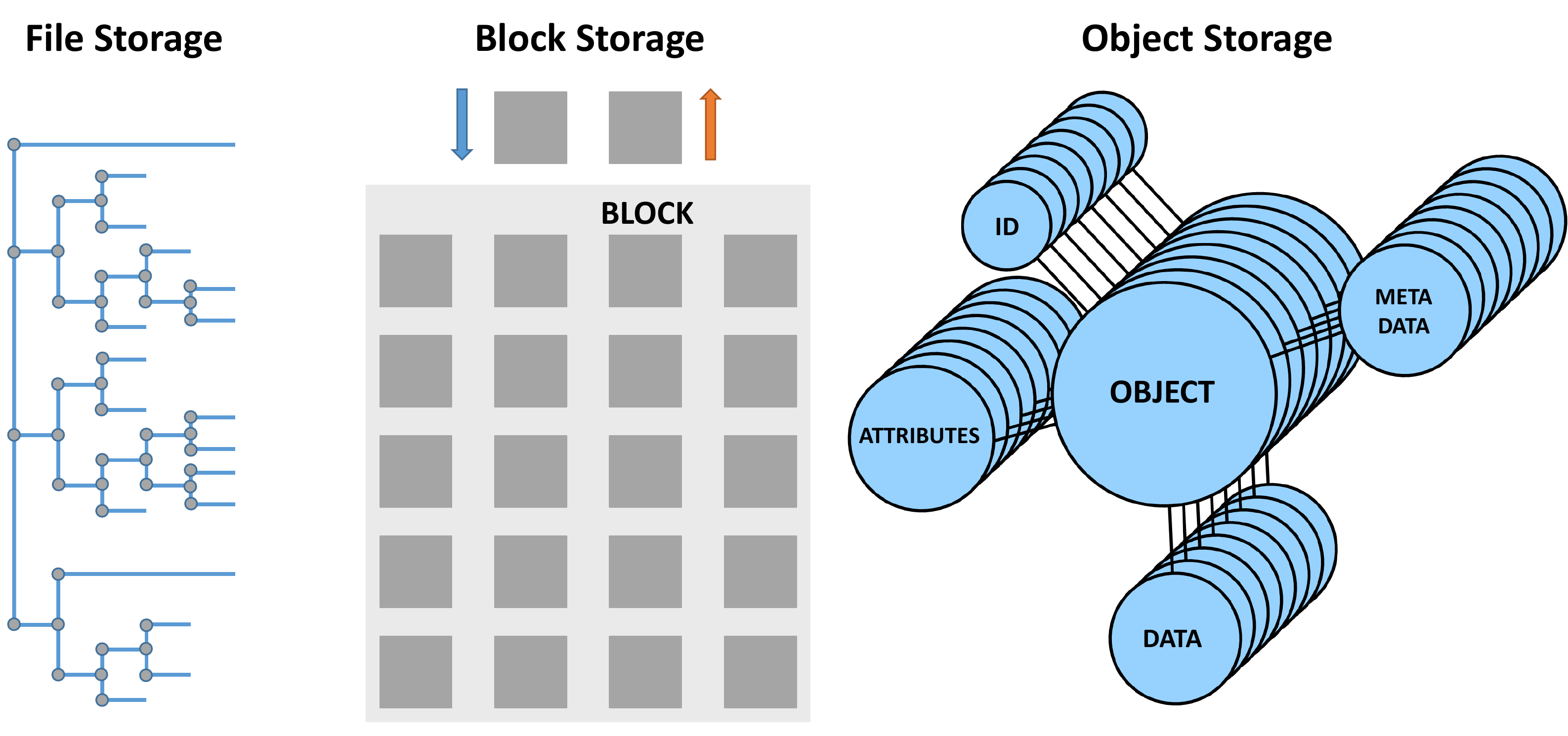}
\caption{Cloud Storage Types } \label{fig:store}
\vspace{-0.4cm }
\end{figure}

Object storage addresses data storage as abstract, discrete units called objects inside a single repository \cite{zheng2013cosbench}. Every object consists of many parts, including the actual data, a globally unique identifier that acts as an address, an expandable amount of metadata along with other relevant attributes (see Fig. \ref{fig:store}). This type of storage has extra protection as usually there are multiple copies in geographically separate regions. In addition, object storage is handling the increasing data growth challenge with architectures that are easily scalable and can be managed by simply adding additional nodes. The flat name space organization of the data, along with the functionality of expandable metadata, are key aspects of this storage service type.

Amazon's native object storage is AWS Simple Storage Service (S3) which offers flexible and low-cost storage. Their storage abstraction is described by the word {\it buckets} and S3 allows an unlimited number of objects (each one limited to 5 TB) per bucket. AWS offers a standard service level with 99.99\% availability on year basis and 11 nines durability. Also, it offers an infrequent service level with 99.9\% availability, the same 11 nines durability, and lower storage costs as a counterweight for high ingress/egress costs. Moreover, Amazon offers AWS Glacier as a form of cold storage designed for data archives and backup functionality. The service provides extremely low prices at the expense of increased latency (four hours required for first byte reception). Finally, AWS supports in-flight and at-rest encryption with different options, including server-side encryption and client-side encryption.

At this point, we should highlight that the AWS S3 is among the services where Amazon engineers have applied formal specifications in an attempt to identify and reduce important design issues \cite{newcombe2015amazon}. Formal specifications are mathematically-based techniques designed to aid the implementation of complex systems and software. Regarding AWS, the TLA+ tool was utilized, which is a formal specifications language based on set theory and discrete math \cite{tla2, tla1}. TLA+ describes the set of all potential execution traces and legal behaviors of a system, along with overall design and correctness properties \cite{newcombe2015amazon}. The tool is used either to examine whether the executable code correctly implements the high-level desired functionality or as an overall aid that helps engineers implement "correct" designs and get a better understanding of them. Finally, formal specifications and TLA+ can reduce errors in code, and discover subtle and significant bugs that are undetected by the traditional extensive design/code reviews and testing.

Azure's object storage offering is called Blob Storage and uses the term {\it containers} instead of buckets. They offer unlimited number of objects per container and up to 500 TB space per storage account. Azure has an alternative view of service levels having the options of:
\begin{itemize}
\item \textit{Locally Redundant Storage (LRS)}, where data are replicated within the same data center (within the account's primary region).
\item \textit{Zone Redundant Storage (ZRS)}, where storage replicated across multiple facilities within the same zone or across two geographical regions.
\item \textit{Geographically Redundant Storage (GRS)}, where data are replicated synchronously locally and then asynchronously to a secondary data center far away.
\item \textit{Read Access Geographically Redundant Storage (RA-GRS)} that adds read-access permissions to the other (secondary) geographic region that is used as a backup data center.
\end{itemize}
Microsoft's cold storage option is Azure \emph{Cool} Blob Storage and unlike the \emph{Hot} option it offers low storage costs with lower availability (see Table \ref{Table:os}).

Google provides a unified object storage option that is the Google Cloud Storage service. Similar to the previous providers, they offer four different service levels:
\begin{itemize}
\item \textit{Multi-Regional Storage} for frequently accessed objects that should be stored geo-redundantly (in at least two geographically separated regions).
\item \textit{Regional Storage}, which offers data stored at a specific region with lower cost.
\item \textit{Nearline Storage}, which offers  data stored for lower cost at the expense of slightly lower availability and minimum storage duration of one month.
\item \textit{Coldline Storage}, as a form of cold storage for infrequently accessed objects designed for archiving and backup functionality.
\end{itemize}
All these storage types provide the same throughput, latency, and high durability of 11 nines. Moreover, all types support creating buckets in locations worldwide, with unlimited object size and storage that can be accessed globally. The differences lie in their availability, storage duration, cost for storage, and access (see Table \ref{Table:os} for the overall summary). 

\begin{figure}[t]
\centering
\includegraphics[width=\columnwidth]{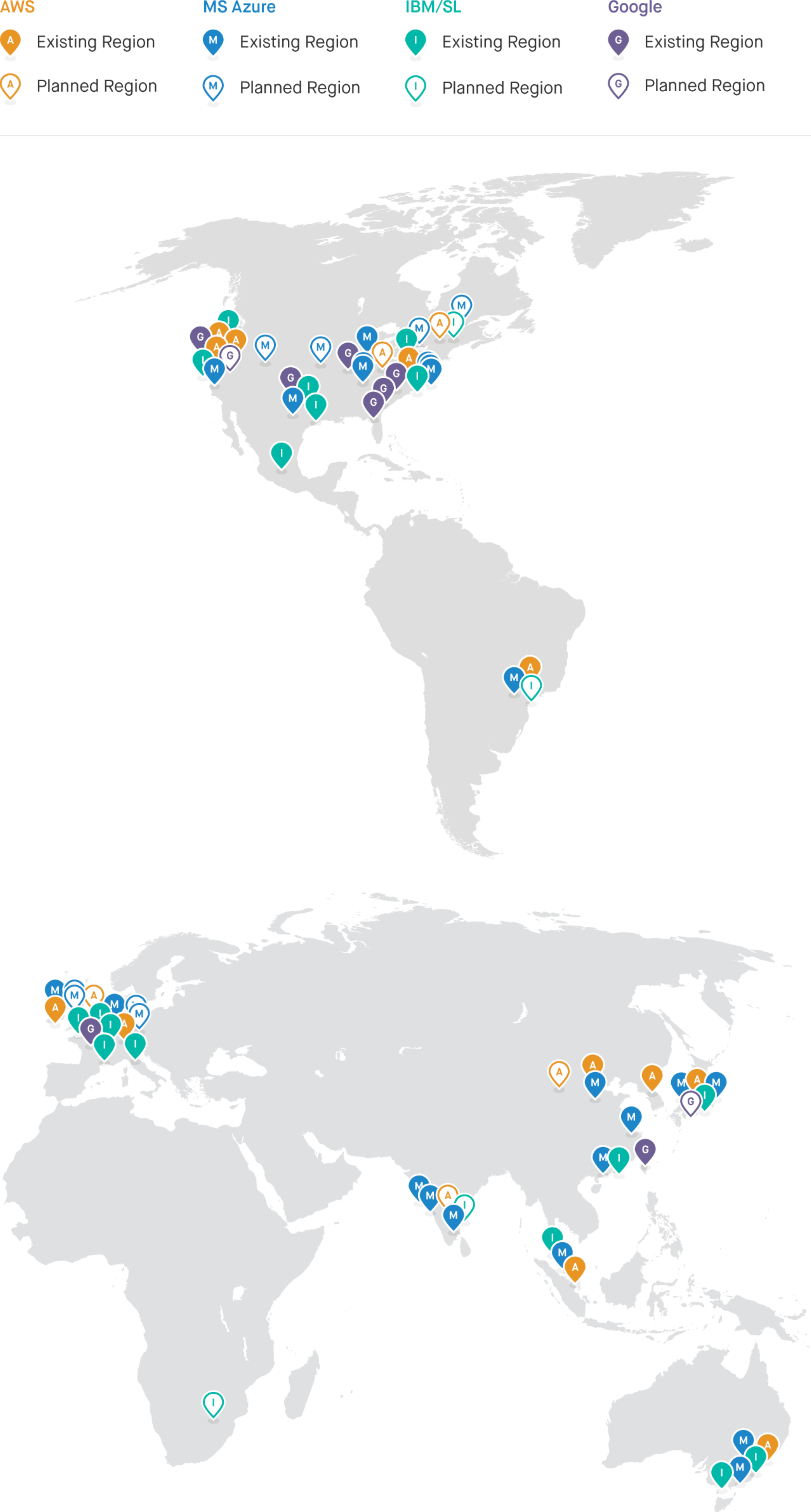}
\caption{Data Centers and Associated Regions of Private Cloud Vendors provided by \cite{map} - Existing and Planned Regions during April of 2016} \label{fig:reg}
\vspace{-0.4cm }
\end{figure}

Finally, IBM's Bluemix Cloud Object Storage service \cite{cloud2016object} is based on the OpenStack Swift platform having a smaller limit per object equal to 5 GB when uploaded thought the API. Further, it provides the ability to create an object in multiple chunks and set a manifest file to automatically store it together with the size reaching 5 TB. Bluemix offers a standard 11 nines durability with regional and cross-regional options. The cross-regional service separates chunks of data to at least three geographical regions focusing on high availability. The regional service stores data in multiple data center facilities in the same region focusing on low-latency. Apart from this classification, IBM offers four different configurations for the Object Storage service, namely:
\begin{itemize}
\item Standard: This service is offered for frequent accessed data and active workloads.
\item Vault: This service is used for infrequently used data, with a 1-month minimum duration, 128 KB minimum object size, targeting archive and backup needs.
\item Cold Vault: This service is also offered for infrequently used data, with a 90-day minimum duration, and a 256 KB minimum object size. Provides lower storage cost with the highest cost for operational requests.
\item Flex: It is used for data that need to be accessed dynamically. Uses an extra dedicated cost model.
\end{itemize}
Table \ref{Table:os} compares
cost, availability, and region support in the public cloud object stores of the four vendors. Fig. \ref{fig:reg} depicts the regions' distribution across the globe during April of 2016.

\subsection{Block Storage}

Block storage provides a more standard storage system configuration by breaking a file into fixed-sized blocks of data and storing them as separate pieces, as shown in Fig. \ref{fig:store}. This is done without a file-folder structure with each block having a unique address. Related services provide a virtualized storage area network with logical volume management provisioning. Each block device can be mounted by a guest operating system the same way as a physical disc. This service provides efficiency as the storage system spreads the smaller data blocks accordingly.

Amazon's offer in this storage category is the AWS Elastic Block Storage (EBS). Volume sizes range from 4 GB up to 16 TB and four volume types are offered as follows: 
\begin{itemize}
    \item Provisioned Input/Output Operations Per Second (IOPS) SSD (io1) Volumes: the highest performance option destined for intensive workloads and offering a maximum of 20,000 IOPS along with 320 MB/s of maximum throughput per volume.
    \item General Purpose SSD (gp2) Volumes that is the standard SSD option delivering a baseline performance starting at 3 IOPS/GB up to 10,000 IOPS and offering 160 MB/s of maximum throughput per volume.
    \item Throughput Optimized HDD (st1) Volumes: the low cost HDD volume option offering a maximum 250 MB/s per TB, and throughput up to 500 MB/s per volume. 
    \item Cold HDD (sc1) Volumes: the option of the lowest cost designed for infrequent workloads and with performance of maximum 80 MB/s per TB, and throughput of maximum 250 MB/s per volume.
\end{itemize}
The service design, similarly to S3, was verified using TLA+ and formal specifications \cite{newcombe2015amazon}.

Microsoft's Azure offers the Managed Disks service providing two categories, namely, Standard Disks and Premium Disks. Azure Standard Storage is a low cost HDD-based offering with volume sizes, from 1GB to 1TB. Maximum throughput per disk is 60 MB/s with 500 maximum IOPS per disk. On the other hand, the Premium Storage offering is a SSD-based service for high-performance, low-latency, IO-intensive workloads providing 128, 512 or 1024 GB disk options. Maximum throughput per disk is 200 MB/s with 5000 maximum IOPS per disk.

In Google's block storage product (Persistent Disk, "PD") volume sizes range from 1GB to 64TB. Google offers two different types: Standard persistent disks and SSD persistent disks. Google is the leader concerning the IOPS offered with 40,000 IOPS for reads and 30,000 for writes to its SSD disks. Maximum throughput per SSD instance is 800 MB/s for reading and 400 MB/s for writing. Regarding Standard persistent disks, Google also offers the highest level of IOPs-per-volume at 3,000 for reads and 15,000 for writes.
For this category, maximum throughput per instance is 180 MB/s for reading and 120 MB/s for writing.

Finally, SoftLayer/Bluemix Block Storage service offers volume sizes of 20GB to 12TB (a smaller size than other cloud providers) with two different volume types: Endurance and Performance. Endurance storage provides less IOPS per GB than Performance storage which is destined for use cases without a high rate of transactions or quick read/write operations.

\subsection{File Storage}

This storage service is the most traditional type, also known as shared filesystem. Data are stored in a file hierarchy (see Fig. \ref{fig:store}), similar to an operating system while multiple clients have the ability to access a single shared folder. The shared filesystem protocols used today are the Network File System (NFS) and the Server Message Block (SMB).

All cloud service providers offer this specific storage type. Amazon provides AWS Elastic File System (EFS),
a service that utilizes EC2 instances and the NFS 4.1 protocol. Microsoft offers Azure File Storage which manages file shares through the SMB 3.0. The stored data are also accessible using a REST API for better integration. IBM also offers a dedicated file storage service, Bluemix's File Storage. Finally, Google relies on its standard Cloud Storage service and to custom Compute Engine instances that can be utilized as dedicated file servers.

\section{Database Services}
\label{section:database}

\begin{table*}[t]
\scriptsize
\renewcommand{\arraystretch}{1.3}
\caption{Database Services}
\label{Table:db}
\centering
\begin{tabular}{||l|c|c|c|c||}
\hline 
\bf{Service Type}   & \bf{Amazon}&  \bf{Microsoft}& \bf{Google}& \bf{IBM}    \\ \hline \hline
Relational Database & AWS RDS & Azure SQL Database & Cloud SQL & dashDB for Transactions SQL Database,\\ 
Management &  AWS Aurora  &  Azure Database for PostgreSQL                  & Cloud Spanner  & IBM DB2 on Cloud, Informix on Cloud \\ \hline
Non Relational Database  & AWS DynamoDB & Azure Cosmos DB & Cloud Datastore &  Cloudant NoSQL DB \\ 
Management  &       &                  & Cloud BigTable & \\ \hline
In-Memory Data Store    & AWS ElastiCache      & Azure RedisCache & App Engine Memcache & Bluemix Redis Cloud\\ 
 & (Redis or Memchached)      & (Redis) &  (Memcached) &  (Redis)\\ \hline
Cloud Extract,    &     &    &   &  \\ 
 Transform, Load (ETL)    &   AWS Data Pipeline  & Azure Data Factory  & Cloud DataPrep  & Bluemix Data Connect \\ \hline
\end{tabular}
\end{table*}

Databases are an essential component of a functioning cloud system, as they have consistently been one of the most popular uses of cloud computing since its first appearance. In practical use, a database combines the functionality of storage and analytics services as users make queries into structured (or unstructured) runtime data to retrieve information. Two main design approaches regarding runtime data systems have been competing during the last decade; relational (SQL) and non relational (NoSQL) database systems \cite{kulkani}.

The relational model had been well established especially after the development of the query language SQL \cite{astrahan1976system,chamberlin1974sequel} in the early 80s, that simplified the manipulation, administration, and general interaction with the database systems. However, since the relational approach faced difficulties in coping with the performance and scalability needs of data-intensive online applications, a family of distributed non-relational systems arrived \cite{decandia2007dynamo, chang2008bigtable}. While, NoSQL databases found initially a large wave of support, along the way they exhibited limitations that mainly pertained to the different (and not fully developed) query languages of each newly introduced database system. This added confusion, along with compatibility difficulties to the existing lack of off-the-shelf operational tools \cite{kulkani}. This development brings us to the present, where SQL interfaces were added on top of Hadoop and Spark \cite{zaharia2010spark}, while  scalable databases fully embracing SQL made their appearance \cite{kulkani, kallman2008h}. Many have followed since this road of aggressive SQL features and syntax integration into recent database systems \cite{cockroach, timescale}. Google's Cloud Spanner release is the most notable example \cite{bacon2017spanner} with their decision to embrace a fully featured SQL system and engine that eliminates many barriers that arise from dealing with different runtime data  systems. Public cloud providers offer all flavours of database systems, as summarized in Table \ref{Table:db}.

\subsection{Relational Database Management Services (RDBMS)}

Relational databases \cite{curino2011relational} rely on tables, columns, rows, or schemas to organize and retrieve data. The AWS Relational Database Service (RDS) provides a number of database types, including Amazon Aurora (Amazon's internal relational database offering), Oracle, PostgreSQL,  MariaDB, MySQL, and Microsoft SQL Server. RDS is an attractive alternative to running an owned instance in EC2 as it takes care of provisioning, patching, and maintenance. RDS consists of various instance types. For example, hardware offerings scale up to 40 vCPUs and 244 GB of memory. In order to implement data storage, and log storage, the service utilizes the Elastic Block Store service. All supported types implement multi-zone replication, with options that include PostgreSQL, MySQL, Aurora, and MariaDB with cross-region read replicas.

Azure's suggestion for the same functionality is SQL Database, a service based on SQL Server. This fully featured cloud database offers active geo-replication and automatic backups with flexible restore capabilities. A second related service offered by Azure is the Stretch Database, which allows on-premise SQL Server instances to save data into Azure SQL Database. Moreover, Azure has recently added the Elastic database pools service, which allows enterprises to optimize costs by running multiple databases utilizing the same resources, thereby maximizing utilization. 

Regarding Google, Cloud SQL is their managed MySQL database solution. Available instance sizes start at 10 GB, and go up to 10 TB with up to 16 vCPUs available, and 104 GB of memory. The service offers automatic zone redundancy as a built-in function, and there is the option of instant restore and backups. Furthermore, apart from the standard Cloud SQL offering Google has recently commercially introduced Cloud Spanner, which is a globally distributed relational database specifically designed for mission-critical cases. Google has been using a version of this database internally \cite{corbett2013spanner} for a long time. The service's advantage is the offering of strong transactional consistency, same as all relational databases, but at the same time, the database can scale horizontally with high availability and global replication \cite{bacon2017spanner}.  Finally, language support includes Python, Go, Java, Node.js along with Java Database Connectivity (JDBC) for compatibility with third-party applications.

IBM supports the same functionality via multiple Bluemix services. DashDB for Transactions SQL Database is a SQL database service optimized for web apps, general and transactional workloads. The service supports all native DB2 drivers, SQL, PureData, .NET, Open Database Connectivity (ODBC), Java Database Connectivity (JDBC), Netezza, and Oracle. It supports either dedicated instances with 8GB RAM, 2 vCPUs, 500 GB of data and logs space, or dedicated bare metal instances with 128GB RAM and 1.4 TB of SSD storage. Finally, IBM's DB2 on Cloud service provides a database on SoftLayer infrastructure.

\subsection{Non Relational Database Management Services}

NoSQL databases do not rely on table structures and use more flexible data models \cite{leavitt2010will} \cite{pokorny2013nosql}. As RDBMS have failed to cover the performance, scalability, and flexibility needs that data-intensive functions require, NoSQL databases have been adopted by mainstream organizations as discussed earlier. NoSQL is used for storing data with flexible structure, which is growing bigger than structured data not fitting anymore the logic of relational databases. Several different varieties of NoSQL databases are used for needs that fall into four main categories: Key-value data stores, Document stores, Wide-column stores and Graph stores. Advantages offered to enterprises by NoSQL databases include scalability, performance, high and global availability and flexible data modeling.

Amazon's NoSQL offering is DynamoDB \cite{decandia2007dynamo, brunozzi2012big} that supports both document and key-value stores for a healthy amount of flexibility. The service supports primary, and secondary indexes on documents with size less than 400 KB. The database reads support eventual consistency, while strong consistency is available if required. This service was the first to be verified, and its design further developed using TLA+ and formal specifications \cite{newcombe2015amazon}. The service is also supported by Amazon's DynamoDB Accelerator (DAX) which is an in-memory, write through cache (fully managed and highly available) able to improve the database's response times up to the level of microseconds by moving away from transitional side cache architectures. Customers do not have to rewrite applications that utilize DynamoDB. DAX can seemingly handle read-through/write-through caching. In addition, the service also includes Autoscaling to support unpredicted database workloads and activity increase. Finally, DynamoDB developers can use the cross-region replication library for their applications to retain database tables synchronization across multiple AWS regions in nearly real time. This functionality is further support by the recently added Global Tables service that enables effortless data replication among regions without update conflicts while retaining high availability.


Cosmos DB \cite{cosmosdb} is Azure's newly introduced high performance, highly distributed NoSQL database. It is designed as a globally distributed database, able to replicate data to any number of regions with a 99.99\% very strong availability. The service supports a variety of APIs,  including JavaScript, MongoDB, DocumentDB, SQL, Gremlin, and Azure Table storage for data query. Cosmos DB supports the use of graph, key-value, and document data in the same service allowing the user to save and query data in the initial form. Finally, Microsoft introduces with this service a number of innovative consistency models (\emph{Bounded Staleness, Session, and Consistent Prefix}) \cite{cosmosconsistency} that go beyond the \emph{Strong}, and \emph{Eventual} consistency models usually offered by other distributed database products. Finally, Azure Cosmos DB was designed following formal specifications and TLA+ \cite{tla1, tla2, cosmosdb}.

Google's alternative is the Cloud Datastore service, that offers strong consistency, and ACID transactions with data being replicated across data centers  located in a single region. Further, Cloud Datastore charges for read/write operations, storage and bandwidth and supports a number of programming languages including Go, Java, JavaScript (Node.js), PHP, Python, and  Ruby. For workload of large scale Google also offers another NoSQL store, the BigTable \cite{chang2008bigtable, ramanathan2011comparison} service. For this service, Google offers an Apache HBase API, which can be integrated with Hadoop or other big data related services. The customers are able to choose between HDD or SDD storage type. While the SSD option provides 10.000 queries per sec (for a single node cluster), it supports one index per table, atomic updates only at low levels, and no support is provisioned for cross-regions replications. Finally, the Cloud Bigtable service charges for 'nodes', storage and bandwidth, and supports Go and Java programming.

Finally, IBM through Bluemix offers Cloudant NoSQL DB. The service is able to scale globally and handles many data types, including text, JSON, and geospatial. Documents can be accessed, saved, or deleted in bulk or individually. IBM handles the management and scalability of the data store allowing clients to focus on their specific application with Java, Node.js, Python, Swift and Mobile Platforms (Android, iOS) being among the supported environments. Their standard plan offers 20GB of free data storage and tiered provisioned throughput capacity, starting at 100 lookups/s, 50 writes/s and 5 queries/s performance. Table \ref{Table:db} summarizes the NoSQL services offered by the four vendors.


\begin{table*}[t]
\scriptsize
\renewcommand{\arraystretch}{1.3}
\caption{Big Data, Analytics, and Data Pipeline Services}
\label{Table:bigdata}
\centering
\begin{tabular}{||l|c|c|c|c||}
\hline 
\bf{Service Type}   & \bf{Amazon}&  \bf{Microsoft}& \bf{Google}& \bf{IBM}    \\ \hline \hline
Hadoop & AWS EMR & Azure  HDInsight & Cloud DataProc & BigInsights for Apache Hadoop\\ \hline
Data Warehousing    & AWS Redshift      &Azure SQL Datawarehouse & BigQuery & dashDB for Analytics\\ \hline
Data Streaming   & AWS Kinesis & Azure Stream Analytics    & Cloud Dataflow &  Streaming Analytics \\ 
                   &              & Azure Event Hub &  &\\\hline
  &      & Storage Queues &  &  \\
Data Queuing  &  AWS Simple Queue Service (SQS)   & Service Bus Queues & Cloud Pub/Sub  & - \\ 
&     & Service Bus Topics/Subscriptions &  &  \\ \hline
\end{tabular}
\end{table*}

\subsection{In-Memory Data Services}

In-Memory databases (IMDS), also known as Main Memory Databases (MMDB) \cite{garcia1992main}, store data entirely in main memory in contrast with the transitional use of persistent disc storage. Since interacting directly in memory is faster, this type of databases provide faster data management functions and lower CPU requirements. Their performance surpasses simple database cashing techniques as they only improve the data retrieval speeds while IMDS architectures speed up also database write operations. This type of service is important for current cloud applications as it can provide distributed in-memory caching without requiring from the customer to take care of scaling and management issues.

Amazon's in-memory cashing offering is AWS ElastiCache. The service offers compatibility with two different open-source engines: Memcached \cite{memcached} and Redis \cite{redis}. Microsoft's high performance caching product is Redis Cache service that provides various options pertaining to available bandwidth, cluster size, availability, and SLAs. On the contrary, Google does not have a focused service for caching but it supports the functionality through its App Engine service with App Engine Memcache. Finally, IBM offers Bluemix Redis Cloud. The service enables the users to run their Redis datastores in IBM's platform with a large variety of pricing plans.

\section{Big Data and Analytics}
\label{section:ai}

Modern Cloud environments provide a nearly ideal engine for exhaustive Big Data processing and analytics \cite{hashem2015rise}. This capability is becoming crucial in the dawn of an era, where the volume, detail, and flow of information generated by various organizations and sources (e.g., IoT, social media) are overwhelmingly increasing \cite{kaisler2013big,khan2014big,cumbley2013big, hilbert2016big}. Cloud infrastructures utilizing their highly distributed architecture and computing capability provide a number of advantages regarding large-scale data handling, including parallel processing, resource virtualization, data storage, minimum operational and maintenance costs, security, and finally, a vast variety of tailor-made data services for the user to choose from \cite{zheng2013service}. In the two following sections, we will discuss the Big Data, Analytics and Data Pipeline services, offered by Amazon, Microsoft, Google, and IBM that span across many categories. Table \ref{Table:bigdata} summarizes the associated offered services.

\subsection{Big Data Managed Cluster-as-a-Service}

Vendors providing Cloud services fetch unlimited benefits by the union of Cloud, MapReduce programming model \cite{dean2008mapreduce}, and Hadoop \cite{cumbley2013big}. Customers get rapidly scalable processing power and storage. Also, the cost of innovation is lower with cost-effective strategies, and on a pay-per-use basis. Thus, businesses can pay for the storage or analytics as they need without making upfront investments (paying for maintaining a system when it is not being used). Additionally, Hadoop cloud platforms offer a variety of instances for all possible uses,  while the clusters handle large volume of data that already exist in cloud storage, thereby minimizing any migration costs.

At the center of Amazon's offerings in this category is AWS Elastic MapReduce (EMR). It is a Hadoop, Spark \cite{zaharia2010spark}, HBase, Flink, and Presto solution that supports an underlying EC2 cluster with the combination of AWS services such as S3 and DynamoDB. EMR is priced hourly for each node offering two different types: Core nodes -acting as both data node and worker node-, and Task nodes -acting solemnly as worker nodes-. The segmentation prevents the loss of Hadoop Distributed File System (HDFS) data and lowers the costs, while the AWS CloudWatch service can be utilized for scaling and monitoring the cluster. Moreover, clusters can be generated and deleted on demand for completion of jobs or they can run for long periods of time. EMR after cluster provisioning monitors slave nodes replacing all unhealthy ones unseemingly. The service also allows direct access to data stored in AWS S3, while language support includes Ruby, Perl, Python, Java, R, C++, and PHP. Finally, Amazon EMR can also run in an Amazon Virtual Private Cloud (VPC) service, where the client can configure networking and security rules.

Azure's alternative is the HDInsight service that supports Apache Hadoop, Spark, HBase, Microsoft R Server, and Kafka in the Azure cloud. HDInsight clusters are configured to store data directly in Azure Blob storage, providing low-latency and increased elasticity in performance and cost choices. Nodes can be added and removed from a running cluster, while the platform supports Java and Python. Moreover, developers can build data processing applications in any environment they prefer. For Windows developers, HDInsight has a rich plugin for Visual Studio that supports the creation of Hive, Pig, and Storm applications. For Linux or Windows developers, HDInsight has plugins for both IntelliJ IDEA and Eclipse, two very popular open-source Java Integrated Development Environment (IDE) platforms. HDInsight also supports PowerShell, Bash, and Windows command inputs to allow for scripting of job work-flows.

Google offers Cloud DataProc in this category. The clusters can start or scale in only 90 second lead time, while the cost depends on the Cloud Compute Engine prices. DataProc service can host MapReduce, Spark, SparkSQL, Hive, and Pig jobs, while the supported language list includes Python, Java, Scala, and R. A Hadoop Distributed File System compliant connector is provided for Cloud Storage to save data before a cluster reboot. On-demand clusters are not supported initially, however, Google Cloud Platform Console (gcloud cli), Cloud Datapro REST API or Google Cloud SDK are all alternatives that provide full control over the cluster to accommodate this option along with advanced management.

Finally, IBM's related product is BigInsights for Apache Hadoop. The main features of the offering include open source Hadoop with a number of tools and capabilities, including Big SQL, BigSheets, Java MapReduce, Big R (R Language integration in Hadoop), and In-Hadoop Analytics. The platform supports Pig, HBase, Hive, and integration with Spark through a separate service Bluemix Apache Spark. Also, provides high integration with the rest Bluemix services offering interfaces for advanced data analytics, social media analytics and extraction/analysis of text.

\subsection{Data Warehouse}

Data warehousing \cite{cuzzocrea2013data, north2017data} supports the functions of efficient data storage to minimize I/O and deliver query results. This is done at high speeds and towards multiple users. They function as central repositories of data from multiple data sources. Information flows into a data warehouse from relational databases, and typically include structured, semi-structured, and unstructured data.

Amazon's product is Redshift \cite{gupta2015amazon} with the most important features being scalability, fast installation, workload management, data compression, a query optimizer, and fault tolerance. Amazon's Redshift is based on a SQL data warehouse and uses Java Database Connectivity (JDBC) and Open Database Connectivity connections (ODBC). Redshift supports integration with other AWS services and  built-in commands that load data and information in parallel to each node from AWS DynamoDB, S3 or EC2. In these services, we can add AWS Kinesis, Elastic MapReduce, Data Pipeline, and Lambda. 

Microsoft Azure's option in this category is SQL Data Warehouse. This service is Azure's first cloud data warehouse which provides SQL capabilities along with the ability to scale within seconds. The architecture is composed of Storage (data are stored in Azure Blob storage), Compute Nodes (the computing power of the service) and Data Movement Service (allows the control compute nodes to communicate, process, and transfer data to all of the nodes). Azure's SQL Data Warehouse customers only have to pay for the query performance they require, which is a differentiating point from other vendor approaches. In addition, Azure enables users to optimize resource and infrastructure utilization while other vendors force customers to delete the existing cluster, backup the existing data and restore them later.

Google offers BigQuery as its low-cost enterprise data warehouse for analytics \cite{ware}. Its model differs the most from our other data warehouse considerations. Firstly, it is serverless. The BigQuery is straightforward to manage projects and datasets in the Google Cloud Platform. Also, it provides quick scaling to petabytes and requires no provision to scale a cluster. Customers can load their data via a streaming API for real-time analytics or transfer data towards other regions of the Google infrastructure.

Finally, IBM's offering is dashDB for Analytics \cite{lightstone2017making}. The service utilizes IBM's BLU Acceleration technology, which ensures the processing data availability in memory. It is more than just a database as it comes with embedded Netezza analytics, linear regression capabilities, decision tree clustering, K-means clustering, IBM Watson, and R support for predictive analytics. The platform is deployed on IBM’s SoftLayer/Bluemix cloud infrastructure with multiple layers of security and encryption if needed. Table \ref{Table:db} summarizes the data warehouse services per vendor.

\section{Data Pipelines}
\label{section:pipe}

\subsection{Streaming Services}

\begin{table}[t]
\scriptsize
\renewcommand{\arraystretch}{1.3}
\caption{Azure Event Hubs vs AWS Kinesis}
\label{Table:str}
\centering
\begin{tabular}{||l|c|c||}
\hline 
   & \bf{AWS Kinesis} &\bf{Azure Stream Analytics}\\ \hline \hline
\bf{Input Capacity} & 1MB/s per Shard & 1MB/s per TU\\  \hline
\bf{Output Capacity} & 2MB/s per Shard & 2MB/s per TU\\  \hline
\bf{Events/s} & 1K & 1K\\  \hline
\bf{Latency} &  10s (Minimum) & 50ms (Average)\\  \hline
\bf{Protocol} &  HTTPS & HTTPS/AMQP 1.0\\  \hline
\bf{Max Message Size} &1MB  & 256KB\\  \hline
\bf{Included Storage} & - & 84GB per TU\\  \hline
\bf{Throughput Flexibility} & Customizable Shards  & Customizable TU’s \\  \hline
\end{tabular}
\end{table}

Cloud streaming services are imperative for applications that require the collection and process of massive amount of data in real-time, with the simultaneous support of multi-tenancy.
Also, such services should provide low-latency, failure tolerance, elasticity, availability, and consistency, while at the same time achieving low-cost for the consumer. Next, we highlight the basic features of each vendor's offering.

Kinesis Streams is the AWS solution for processing information pipelines. Enterprises can transfer data in real time to a Kinesis stream for processing using the Connector Library and Kinesis Client Library. The service uses the shards as base throughput unit, which signifies a capacity of 1MB/sec data input and 2MB/sec data output. As data unit Kinesis uses the {\it record} consisting of a partition key, a unique identifier and a data blob of 1 MB maximum capacity. The initial number of shards is customizable (without upper limit) supporting up to 1,000 put-records requests per second (single call that writes multiple data records) and up to 5 read transactions per second. 

For processing real-time data streams Azure has built Stream Analytics. The service has the ability to process Blob storage data or information streamed through Event Hubs/IoT Hub. A SQL-based language is utilized to perform queries and can also support the Azure Machine Learning service. Azure Event Hubs stream capacity is described by a throughput unit that includes up to 1MB/sec (or 1000 events/sec) ingress (inbound data), and up to 2MB/sec egress (outbound data). Event publishing can be achieved through HTTPS or alternatively AMQP 1.0 with an event instance capacity limit of 256 KB. A side by side comparison of these two provided streaming services can be found in Table \ref{Table:str}.

In Google's Cloud Platform the Cloud Dataflow service can be used to build data processing pipelines. Google's approach differs from AWS and Azure. The aforementioned services offer a model that delegates processing to adjacent services such as Hadoop. Google's Cloud Dataflow, on the other hand, supports a fully programmable (Python, Java) framework and a distributed computing platform. The service also supports both batch and streaming workers with their number being pre-defined when the service is created. Batch workers have the option to auto-scale on demand/load. Currently, a single user is allowed to make 5000 requests per second with up to 25 simultaneous Dataflow jobs.

IBM's answer is the Streaming Analytics service that consists of a programming language, an API, an IDE for applications, and a runtime system that can run the applications on a single or distributed set of resources. Streams processing applications can be developed in multiple supported languages,  including Java, Python, and Scala. Their standard plan offers 4-core virtual server nodes with 12GB of RAM and 1Gbits/second Network, while in the premium offering each node is a 16-core virtual server with 256GB of RAM, 2TB of disk and unlimited public bandwidth at 100 Mbps.

\begin{table*}[t]
\scriptsize
\renewcommand{\arraystretch}{1.3}
\caption{Machine Learning and Artificial Intelligence Services}
\label{Table:ml}
\centering
\begin{tabular}{||l|c|c|c|c||}
\hline 
\bf{Service Type}   & \bf{Amazon}&  \bf{Microsoft}& \bf{Google}& \bf{IBM}    \\ \hline \hline
Machine Learning & AWS Machine Learning  & Azure Machine Learning &Cloud Machine Learning & IBM Watson Machine Learning\\
& & &Engine&\\ \hline
Language Processing \&    & AWS Lex      & Azure Cognitive Services & Cloud Natural Language API & Natural Language Understanding\\
Speech Recognition AIs & AWS Polly &  & Cloud Speech API  & Speech to Text/ Text to Speech\\ 
   &       &  &   &Conversation, NL Classifier  \\ \hline
Image Recognition AI   & AWS Rekognition      & Azure Cognitive Services & Cloud Vision API  & Visual Recognition \\ \hline
\end{tabular}
\end{table*}

\subsection{Queuing Services}
Message queuing API's are important cloud federation building blocks. Such services are used to couple cloud application components and move messages among highly distributed and diverse environments with high reliability \cite{yoon2012interactive}. 

Amazon's product is AWS Simple Queue Service (SQS) that supports queues for storing messages as they move between different cloud components. Their API is supported by many programming languages,  including Java, Ruby, Python, .NET, PHP, and Java Script/Node.js. The service also offers two variations of queues:

\begin{itemize}
    \item {\it Standard Queues}: Allow a large number of transactions per second, while WAS claims a single-time message delivery guarantee with a best-effort policy for ordering.
    \item {\it First-In First-Out (FIFO) Queues}: Guarantee message delivery once and strictly preserves sent and received order. The service allows 300 transactions per second and per action. They are used for applications where messaging order is critical. 
\end{itemize}
SQS message size is 256 KB of text data (XML, JSON or unformated), while regarding queue sizes there is a 20000 limit for FIFO and 120000 for standard queues. Finally, typical latencies regarding queuing actions vary from 10 to low hundreds of milliseconds.

The Microsoft equivalent of AWS SQS consists of two services: Storage Queues and Service Bus Queues. The first is mainly part of the Azure Storage family and provides reliable message exchange between services. Storage Queues offers no guarantees regarding ordering and an at-least-once delivery policy. Maximum queue size reaches up to 500 TB with unlimited number of queues, while the maximum size of a message is 64 KB. On the other hand, Service Bus Queues are part of Azure's messaging infrastructure offering FIFO ordering guarantee with an addition of at-most-once delivery policy. The maximum number of queues is limited to 10000 with 1 to 80 GB maximum queue size and 256 KB/1 MB max message size. The aforementioned two services support both REST over HTTPS as their management and runtime protocol and also APIs for .NET, C++, Java, PHP, and Node.js. Further, Azure offers in addition to Service Bus Queues, where each message is processed by a single entity, the Service Bus Topics, and Subscriptions where a message is broadcasted to multiple resources in a publish/subscribe fashion. This subservice is used to scale the queuing functionalities to many recipients, as it resembles a virtual queue where messages are sent to a specific topic and are received to one or more associated subscriptions.

On the contrary, Google has not a specific queuing service. The functionality is supported by its cloud PUB/SUB service, an engine that allows message exchange between individual entities. This is achieved virtually using a single topic and subscription logic, where an \textit{at-least-once} delivery policy is implemented and a FIFO (in-order) guarantee is not supported. Client libraries include GO, C\#, Node.js, Java, PHP, Python, and Ruby. Finally, IBM does not support such functionality through a dedicated cloud service inside Bluemix, but only as part of the general ecosystem.

\section{Machine Learning and Artificial Intelligence}
\label{section:ma}

Machine Learning and Artificial Intelligence (AI) technologies provide tremendous opportunities for advances that will improve human lives in a vast collection of sectors that include healthcare, education transportation, public safety, and entertainment. Inevitably, as progress in AI is relying on machine learning (ML), big data, communications, and analytics, cloud computing and the related resources are becoming the number one platform able to deliver AI services. Moreover, as spending on cognitive and AI systems is estimated to overcome the threshold of 45 billion dollars by 2020 \cite{idcAI}, all major cloud providers work to build, grow and properly equip their services to meet all the diverse application needs. Since in this specific category the offered services are constantly being updated, we will list the different services offered focusing mainly in the different use cases. Table \ref{Table:ml} summarizes the associated offered services.

Amazon offers AWS Lex as its language processing and speech recognition AI. The service provides conversational interfaces for text and voice (the engine is also behind Amazon's Echo product). Currently, the service supports the English language with 15 seconds speech input of either Linear Pulse Code Modulation (LPCM) or Opus \cite{opus} format. For text-to-speech conversion, Amazon provides AWS Polly. This service adds a spoken response to applications with over 17 unique supported languages and MP3 or raw PCM audio output formats. Further, for image processing applications,  Amazon offers AWS Rekognition, which is based on deep learning architectures. Currently, image input formats include JPEG and PNG with sizes up to 15MB (through the S3 storage service) and 5 MB if provided directly. Also, Amazon Rekognitionm supports a variety of image labels used to extract common categories and supports facial analysis, comparison, and recognition with the ability to detect 12 different facial attributes. Finally, the general ML offering of Amazon is the AWS Machine Learning. This service is able to train models from datasets of size up to 100 GB and can yield real-time predictions within 100 ms. All Amazon AI services are supported by the same language APIs that include Java, .NET, Node.js, PHP, Python, Ruby, Go, C++, Android, and iOS.

Microsoft follows a differing strategy by offering the Azure Cognitive Services, which includes APIs for different AI functionalities. Regarding language processing, Azure offers the Language Understanding Intelligent Service (LUIS) an application with HTTP endpoints that provide written language understanding capabilities. Azure also offers the Translator Text API and the Text Analytics API (V 2.0) that provides functions including key phrase extraction (for 5 languages) and sentiment analysis (for 15 languages). The data limits include 10 KB of single documents, 1 MB maximum size of an entire job, and 1000 maximum number of documents per job. For processing spoken language, Microsoft offers the Translator Speech API, the Custom Speech Service, which supports speech-to-text transcriptions, and the Speaker Recognition API for speaker identification. Moreover, Azure offers face recognition capabilities through Emotion API and Face API. General image processing needs are supported via the Computer Vision API. Input image formats include JPEG, GIF, BMP, and PNG with maximum size 4 MB and at least 50 $\times$ 50 pixels dimensions. Language supports C\#, Java, PHP, JavaScript, Python, and Ruby. Finally, Azure offers the Machine Learning service for more general ML applications with dataset support of up to 10 GB (multiple inputs). Scripting modules include the support of SQL, R, and Python, while for new custom modules the customers can only use R.

Google also offers a number of different products for cloud-based machine learning applications. The general ML offering is the Cloud Machine Learning Engine. For text analysis purposes, the Cloud Natural Language API is offered packing label, syntax analysis, and sentiment extraction. The API currently supports 9 different languages, while the Cloud Client Libraries for developers working with the Natural Language API include Java, PHP, Ruby, Python, Node.js, C\#, and Go. Image processing is handled by the Cloud Vision API that can provide among others face, logo, landmark and any custom content detection on the input image. Supported formats include JPEG, PNG, GIF, BMP, WEBP, and ICO, while their size should be under 4 MB. Regarding the image sizing 640 $\times$ 480 pixels is the standard, while recommended sizes for different types of requested jobs are 1600 $\times$ 1200 pixels (face detection), 640 $\times$ 480 pixels (landmark, logo, label detection) and 1024 $\times$ 768 pixels (text detection). A separation point for Google's ML deployments is the use of custom Tensor Processing Units (TPUs) \cite{jouppi2017datacenter} that accelerate neural network computations. TPUs are custom Application-specific Integrated Circuits (ASICs) built specifically for machine learning and the TensorFlow technology, which is an open source software library developed by the Google Brain Team \cite{abadi2016tensorflow, abadi2016tensorflow2}. TensorFlow allows numerical computation using data flow graphs, where mathematical operations are represented by specific nodes and tensors ($\equiv$ m-dimentional data vectors or arrays) are represented by the graph edges. TensorFlow is used to basically program TPUs that can be combined with each other (forming ML supercomputers) or with other Google hardware such as CPUS or GPUs. The technology provides significant improvements regarding model training times and provide the opportunity for the customer to integrate ML accelerations directly to their existing product.

Regarding IBM, the Watson Machine Learning service is the main powerhouse behind cognitive application development. The service is composed of a set of REST APIs called from any programming language. It allows integration with the IBM SPSS Modeler, which is a data mining and text analysis workbench that requires little or no programming to operate. Service offers a totally free plan with the ability to deploy up to 5 models with 5000 predictions per month and a 5 hour per month restriction in compute time. On the other hand, the fully optimized professional plan offers 2000000 predictions and 1000 compute hours with extra billing for any extra hour or any extra 1000 predictions. For language processing and text reckognition applications, IBM Bluemix offers a number of dedicated services. The Natural Language Classifier service understands and processes text that can be provided in CSV format, UTF-8 encoded and with maximum 15000 rows or 1024 characters. The service supports 9 different languages and can be handled currently using Node.js, Python or Java APIs. A related service is the Natural Language Understanding that can be used to analyze semantic features of text input such as emotions, labels, sentiment, or other entities in 11 languages. The Bluemix Speech to Text service converts human voice from 8 languages to written text with the current version supporting Java and Node.js libraries. The reverse service Text to Speech is also offered supporting Pulse-Code Modulation (PCM), MP3, Opus or Vorbis codec, Waveform Audio (WAV), Free Lossless Audio Codec (FLAC), Web Media (WebM) format, or basic audio. The interfaces available include HTTP REST API and a WebSocket interface to synthesize text. Bluemix Watson Conversation service deploys a natural language interface in the customers application with 13 supported languages and the ability to extract from the audio input: purposes/goals (Intents), classes of objects/data types (Entities), and dialog. Finally, Watson Visual Recognition service is used for image analysis. The service is able to detect facial characteristics along with specific themed tags in each image from multiple categories. It can accept up to 10000 per .zip file (or 100 MB) with minimum image size equal to 32 $\times$ 32 pixels.

\section{Networking and Content Delivery}
\label{section:net}

\begin{figure}[t]
\centering
\includegraphics[width=\columnwidth]{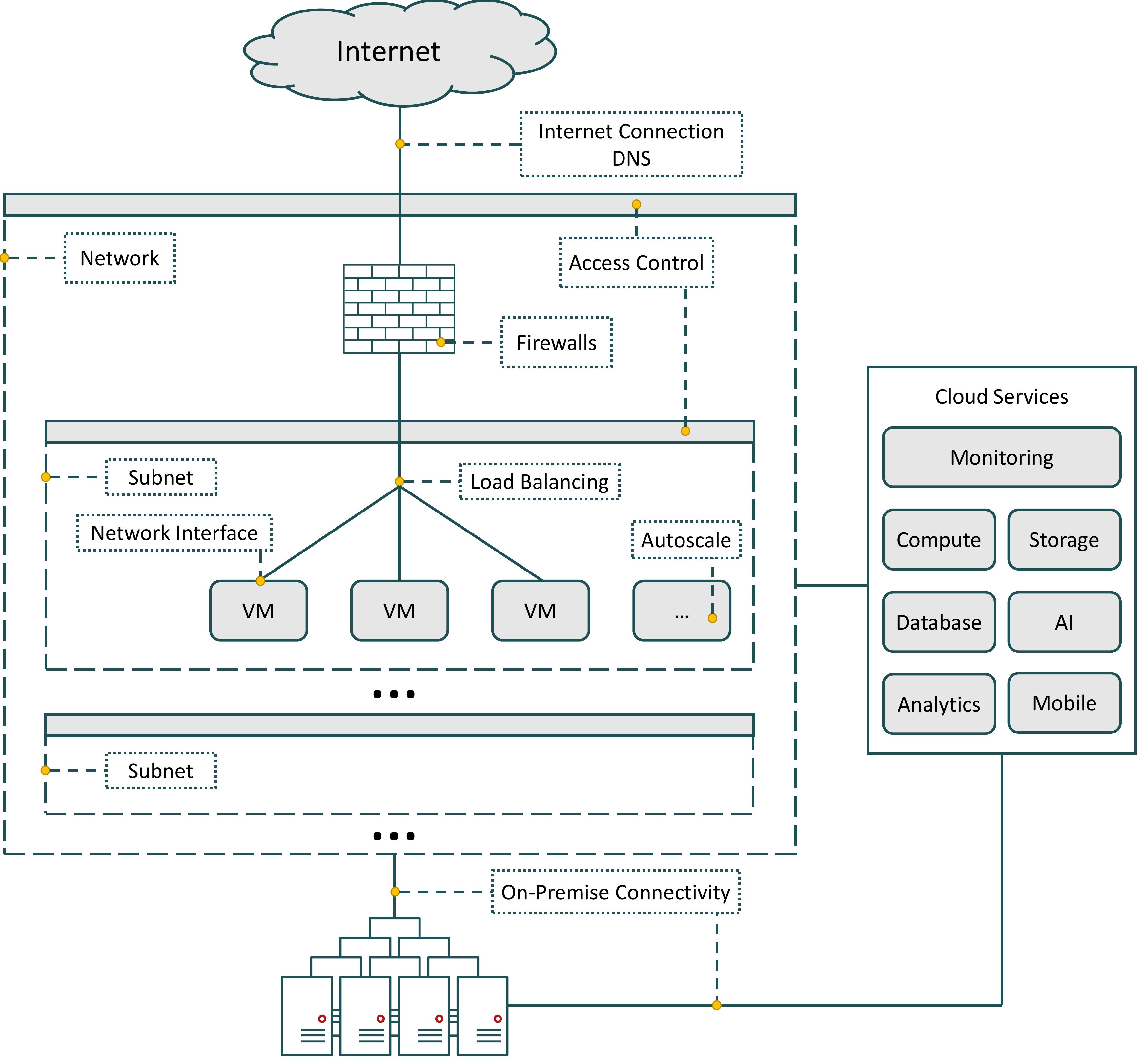}
\caption{Networked Cloud Application} \label{fig:app}
\vspace{-0.4cm }
\end{figure}

A traditional application running on the cloud is shown in Fig. \ref{fig:app}. Networking services provide the essential connectivity between the cloud applications, stored data, and the rest components of the cloud infrastructure. In this section, we will discuss the main networking services offered by the four cloud vendors, while Table \ref{Table:ns} summarizes them.

\subsection{Virtual Networking}

Virtual Networking is a vital type of service that provides a virtual network inside the cloud infrastracture of any vendor. Offered components include subnets, internet gateway, virtual private gateway, Network Address Translation (NAT) gateway, routers, peering connections, customer gateway, and hardware Virtual Private Network (VPN) connections. These services also provides VNet isolation, on-premises connections, and network traffic routing and filtering. Amazon's offering is called Virtual Private Cloud (VPC), Microsoft's product is Azure Virtual Network service, Google has the Cloud Virtual Private Cloud (VPC), and finally IBM's offering is Network Appliances.

\begin{table*}[h]
\scriptsize
\renewcommand{\arraystretch}{1.3}
\caption{Networking Services}
\label{Table:ns}
\centering
\begin{tabular}{||l|c|c|c|c||}
\hline 
\bf{Networking Services}   & \bf{Amazon}&  \bf{Microsoft}& \bf{Google}& \bf{IBM}    \\ \hline \hline
Virtual Networking & AWS VPC & Azure VNet &  Cloud Virtual Private Cloud  & Network Appliances\\  \hline
DNS Services  & AWS Route 53 & Azure DNS & Cloud DNS &  Domain Name Service  \\  \hline
Private Connectivity    & AWS Direct Connect     & Azure Express Route & Cloud InterConnect & Direct Link \\ \hline
Content Delivery Network    & AWS CloudFront      & Azure CDN & Cloud CDN & Content Distribution Network \\ \hline
Load Balancing   & AWS Elastic Load Balancing  & Azure Load Balancer  & Cloud Load Balancing & Bluemix Load Balancer \\ 
   & Classic \& Application Load Balancing & Azure Application Gateway  &  &   \\ \hline
\end{tabular}
\end{table*}

\subsection{DNS Services}

Domain Name System (DNS), is the part of the Internet that allows site access using host-names. The process involves a translation of a host-name to an IP address via querying a DNS server with assigned responsibility for that hostname. 
All cloud vendors provide such authoritative DNS services that allows users to manage their public DNS names. 

AWS Route 53 is Amazon's reliable and cost-effective Domain Name System (DNS) web service that translates domain names into numeric IP addresses. It is authoritative and effectively connects client requests to systems running in AWS – AWS EC2 instances, ELB instances, or AWS S3 buckets– and can also be used to route users outside of AWS. Route 53 allows an enterprise to manage traffic around the world through a variety of routing types, including Geo DNS, Weighted Round Robin, and Latency Based Routing. Organizations can also use Amazon Route 53 monitor their application and endpoints' health. In addition, it is used for routing traffic towards healthy endpoints.

Microsoft's offering is Azure DNS, also an authoritative DNS service that allows organizations to manage their DNS names. Being an Azure service, it allows network administrators to benefit from all the access controls, auditing, and billing features. Azure DNS is the fastest service but also the cheaper comparing to the AWS and Google offerings.

Google's solution is Cloud DNS. It is characterized by low-latency, availability, and low-cost in making applications or services available to users. It provides 100\% availability and low-latency with automatic scaling.

Finally, IBM offers the Bluemix Domain Name Service with the primary benefit over another DNS management services being that the client has a central, reliable location in which all of the data are stored. As an additional service, Bluemix offers secondary DNS zones to the customers free of charge, allowing users to back up their primary DNS records in the event of data loss or node failure.

\subsection{Private Connectivity}

All the major cloud providers offer the possibility to connect to them directly, and not over the Internet. This solution significantly aids the cloud infrastructure in terms of:
\begin{itemize}
\item Bandwidth: getting guaranteed bandwidth to the cloud and its applications.
\item Latency: having an explicit maximum latency on each connection.
\item Privacy/security: by avoiding exposure of the traffic on the Internet.
\end{itemize}
All vendor offerings in this category -Amazon's AWS Direct Connect, Microsoft's Azure ExpressRoute, Google's Cloud InterConnect and IBM's Direct Link- support connection of private networks to cloud networks over a leased line rather than using the Internet. Regarding the differences among them, AWS Direct Connect is a IEEE 802.1q VLAN (layer 2) based service. There is an hourly charge for the port (that varies by the port speed), and also per GB egress charges that vary by location (ingress is free, just like on the Internet). Azure's ExpressRoute is a BGP (layer 3) based service, and it also charges by port speed, but the price is monthly, without any further ingress/egress charges.
An interesting recent addition is ExpressRoute Premium, which enables a single connection across Microsoft’s private network into many regions rather than having point-to-point connections into each region. As for Google's InterConnect, it is a BGP (layer 3) based service. The connection itself is free, with no port or per hour charges. Egress is charged per GB, and varies by region. Finally, the Bluemix Direct Link offering supports secure layer 3 connectivity between customer remote network environments and associated computing resources. Direct Link is essentially an alternative to the traditional site to site VPN solutions for users that require more consistent, and higher throughput between remote networks and their cloud environments.

\subsection{Content Delivery Network}

A Content Delivery Network (CDN) is a network of geographically distributed servers and data centers that provides multiple types of content (web objects, media files, documents, live-streaming content, etc.) to end-users \cite{pathan2007taxonomy}. Their architecture replicates content from an origin server to cache servers located in different regions around the world aiming to reduce latency in delivery, reduce site load times, provide maximum availability, eliminate geographical barriers, and provide better management during traffic surges. Since CDNs are able to scale delivery on a global scale, many companies favor them and cloud vendors already have the distributed infrastructure to support them.

Amazon's service is CloudFront, which is deeply integrated with the rest of AWS services such as S3, EC2, Route 53, Lambda or Elastic Load Balancing. The infrastructure spans across 16 geographic regions (offering 44 availability zones) and imminent plans for expansion to a global network of 82 edge locations. CloudFront supports all files able to be served over HTTP/HTTPS, while regarding streaming protocols AWS supports Adobe’s Real Time Messaging Protocol (RTMP), Adobe’s HTTP Dynamic Streaming (HDS), Apple’s HTTP Live Streaming (HLS), and Microsoft’s Smooth Streaming. Regarding the maximum single file size that can be delivered, it is set at 20 GB. Microsoft offers the Azure Content Delivery Network (CDN) service to cache web content globally. Over 45 unique CDN edge points are deployed with management APIs that include REST, .NET, Node.js, or PowerShell. Google offers its own Cloud CDN service that supports also HTTP/2 and HTTPS requests and operates caches at over 80 locations around the world. Finally, IBM offers the Bluemix Content Distribution Network service with 24 nodes globally and support of various encoding formats for caching media with  DivX, H.264, Silverlight, QuickTime, MP3, HTML, TXT, PDF, GIF, and JPG being some cases in point. 

At this point, we should underline that all four CDN services offer features that pertain to:
\begin{itemize}
    \item Performance optimization 
    \item Security (HTTPS support, geo-filtering, token authentication DDoS attack protection)
    \item Logging, Reporting, \& Analytics.
\end{itemize}

\subsection{Load Balancing}

Load balancing is one of the key components in the cloud architecture. It is a process that ensures the distribution of workloads in excess, evenly, and aims to optimally balance their load among the available resources (compute nodes, memory, network, or storage) \cite{randles2010comparative, al2012survey, xu2017survey}. The focus lies with optimal resource utilization towards better throughput, smaller response or migration times, optimal scalability, and overall system performance. As it is one of the most basic services in a cloud environment, all vendors offer related products to distribute incoming traffic towards their settings.

Amazon offers the AWS Elastic Load Balancing service, providing automatic scaling, security, and high availability. AWS separates the service into two sub-services:
\begin{itemize}
    \item {\it Classic Load Balancer}: Used for cases that require traffic load balancing across EC2 instances. Traffic is routed following network level or limited application-based criteria. The service supports application that use TCP/SSL (Layer 4 load balancing) or HTTP/HTTPS protocols.
    \item {\it Application Load Balancer}: Used for cases that require traffic load balancing across ports, microservices, or docker-based services across the same EC2 instance. Traffic is routed following application-based criteria. The service supports application that use HTTP/HTTPS protocols and WebSockets. The cost for this category is approximately 10\% lower, while it additionally supports load balancing using IP address as targets.  
\end{itemize}

Microsoft's product is Azure Load Balancer, supporting Layer 4 (TCP, UDP) load balancing. Possible configurations include balancing traffic from the Internet to VMs, or balancing traffic from a virtual network and between VMs (either on-premise or cross-regions). For the distribution purposes, Azure utilizes a hash-based algorithm that creates tuples consisting of the destination IP-Port pair, the source IP-Pair, along with a setting related with the used protocol. Apart from this service, Azure offers the Application Gateway that implements Layer 7 (application Layer) balancing with HTTP/HTTPS and WebSockets support. It can be used to route traffic towards any internal or public IP address, VM, or other cloud service. Finally,  the offered Traffic Manager service balances traffic towards endpoints across the world on a DNS level.

Google's Cloud Platform has built the Cloud Load Balancing service to support the functionality, similar to the previous vendors. The service can support either HTTP/S load balancing, or Network load balancing. The first category acts on a HTTP/2-HTTP/1.1 translation layer with cross-region traffic balancing based on IP address, and also content-based balancing based on the incoming URL. The Network load balancing supports the functionality using as criteria information such as address port, or protocol type. It is used to balance UDP, TCP, or SSL traffic (SSL/TCP Proxy load balancing), while it is the service that distributes load within a given region.

Finally, IBM offers the Bluemix Load
Balancer. The service is built to operate on a Layer 4 level for applications that support HTTP/HTTPS and TCP. It can be applied to both virtual and bare metal servers that IBM offers, while the service utilizes an algorithm based on round-robin, and weighted round-robin. In addition, the Local Load Balancer implements internal traffic balancing, while IBM also supports the Citrix NetScaler, a service used for high-performance needs and additionally offers DNS-based traffic balancing. NetScaler can also implement DNS-based load balancing.

\section{Additional Services}
\label{section:rest}

Until now, we have performed a taxonomy of cloud services that can be characterized as key-offerings for any competitive vendor in the industry. However, there is a significant number of services offered at the various cloud environments that handle a number of critical issues such as security \cite{xiao2013security, ryan2013cloud, ali2015security, almorsy2016analysis}, identity authentication \cite{habiba2014cloud}, application development, monitoring \cite{aceto2013cloud, ward2014observing, fatema2014survey}, cloud management, Internet of Things (IoT) support. On account of completeness, this section provides a brief mapping of services that pertain to the aforementioned general categories as implemented and offered by the four vendors we have been examining. We do not discuss the details about these services, as they are either recently added and therefore prone to imminent adjustments, or offer the same functionality to the customers (ignoring cost factors). To that end, Table \ref{Table:other} describes a number of important other services, as offered by cloud service providers.

\begin{table*}[h]
\scriptsize
\renewcommand{\arraystretch}{1.3}
\caption{Additional Services}
\label{Table:other}
\centering
\begin{tabular}{||l|c|c|c|c||}
\hline 
\bf{Service Description}   & \bf{Amazon}&  \bf{Microsoft}& \bf{Google}& \bf{IBM}    \\ \hline \hline
Identity \& Access Management & AWS Identity \& Access  &Azure Active Directory   &  Cloud IAM   & Bluemix App ID, Passport  \\  
&  Management (IAM)   &  &    &   \\  \hline

Security Assessment & AWS Inspector  &Azure Security Center   & Cloud Security Scanner &  Adaptive Security Manager    \\  \hline

Hardware Based Security & AWS Hardware Security Module (HSM)  & Azure Key Vault  & Cloud Key Management   & Key Protect    \\  
Secure Key Management & AWS Key Management Service (KMS) &   & Service (KMS)  &     \\  \hline

Directory Services & AWS Directory Service  & Azure Active Directory  & Cloud Identity-Aware Proxy &  Single Sign On \\  
Single \& Multi-Factor & Multi-Factor Authentication (MFA)   & Multi-Factor Authentication  & Security Key Enforcement   & Adaptive Security Manager \\  
Authentication &      &  Authentication  &    &   \\  \hline

Network Security \& Firewall  & AWS Web Application Firewall (WAF)     & Azure WAF  & Cloud Security Scanner   &  Hardware Firewall \\

 & AWS Shield    & Azure Network Watcher  &    &   \\  \hline \hline  \hline

Management Tools & AWS Management Console  & Azure Management Console  & Cloud Console, Shell   & Bluemix Catalog  \\ 
& AWS Command Line Interface  & Azure CLI, PowerShell  & Cloud  Deployment Manager  & Infrastructure Controls \\ \hline

Monitoring, Logging,  & AWS CloudTrail    & Azure Log Analytics   &  Cloud Stackdriver  & Infrastructure Monitoring  \\  

Error Reporting & AWS CloudWatch    & Azure Application Insights  & Monitoring, Logging,   & Infrastructure Reporting \\ 
 &     & Azure Portal   & Error Reporting, Trace   &   \\ \hline \hline  \hline

Software Development & AWS Code Star, CodeBuild, Cloud9  & Visual Studio Team Services   & Cloud SDK, Cloud Tools   & Foundry, DevOps Insights   \\ \

 & CodeCommit, CodeDeploy,    & DevTest Labs  & Cloud Source Repositories   &  Continuous Delivery \\ 
 
  &  CodePipeline, AWS X-Ray   & Application Insights  & Error Reporting, Trace   &  Globalization Pipeline \\ \hline 
  
Deployment Templates & AWS CloudFormation    & Azure API Management   & Cloud Resource Manager  & Boilerplates  \\ \hline 

API Management  & AWS API Gateway    & Azure Resource Manager   & Cloud Endpoints  & API Connect  \\ \hline  \hline \hline

Mobile App Development  & AWS Mobile Hub    & Azure Mobile Apps   &Cloud Mobile App   &  MobileFirst Services, Swift \\ 

  & AWS Mobile SDK    & Azure Mobile SDK   & Cloud App Engine  & Mobile Foundation  \\  \hline

Mobile App Testing \& & AWS Device Farm     & Azure DevTest Labs    & Cloud Test Lab   &  Mobile Analytics \\
Analytics  &AWS Mobile Analytics     & Xamarin Test Cloud    & Firebase Analytics   & Bitbar Testing  \\ 
   &   & Hockey App    &   & Kinetise   \\ \hline \hline \hline
   IoT Platform \&    & AWS IoT Platform       & Azure IoT Platform & & \\ 
Development Solutions  & AWS Greengrass & Azure IoT Edge & Cloud IoT Core   & Internet of Things Platform   \\ 
      &     & Azure IoT SDK  & &\\ \hline

\end{tabular}
\end{table*}

\subsection{Internet of Things}

Closely connected with cloud-based services is the evolution of the Internet of Things paradigm \cite{gubbi2013internet, munir2017ifciot} with all the associated streams of new data loads. Modern fully-packed IoT platforms are an obvious extension for hyper-scale cloud vendors \cite{botta2016integration}. Relying on the underlying compute, database, network, and security infrastructure IoT offerings should cover a variety of services, including secure data handling \cite{mendez2017internet} and bi-directional communication between edge devices with the addition of data processing \cite{pflanzner2016survey}.

Amazon offers the AWS IoT Platform for supporting complete IoT solutions. Device communications (publish and receive messages) are handled over HTTP, Message Queue Telemetry Transport (MQTT) \cite{mqttProtocol} or MQTT over WebSockets. The service also provides SDKs for Embedded C, JavaScript, Python, iOS, and Android, while messages are processed in 512 byte blocks (= single message up to max 128 KB). Moreover, it offers a strong security design and the service, which is highly integrated with Amazon's internal authentication engine - Identity and Access Management (IAM). In addition, this year Amazon launched AWS Greengrass, a software service that acts as the local gateway to IoT devices, tackling primary routing, data caching, and message processing. Amazon is focusing on edge computing architectures making the framework lightweight enough to be supported by ARM-based System on Chip (SoC) devices. In addition, the service offers M2M exposure through MQTT endpoints, while developers can utilize custom Lambda functions.

Azure's offering in this category consists of the IoT Platform and the IoT Hub. The service supports AMQP, MQTT, and HTTP protocols while supported SDKs include JavaScript, .NET, Java, Python, and C. At the same time, IoT Hub keeps track of devices via a dedicated registry and provides reliable communication between them. Blob Storage handles received data for archiving or offline processing. There is also the alternative of transferring data to an Event Hub instance for real-time processing, monitoring or diagnostics. Messages are sent in 4 KB blocks, while the service has 4-tiers which can support up to 300,000,000 messages per day. Recently, Azure is also focusing on edge computing deployments, by launching the IoT Edge, a software able to run on both Linux and Windows also supporting x86 and ARM architectures. Azure IoT Edge enables the local deployment of Azure services with the language support of Java, C, C\#, Python, and Node.js. Communications are implemented through MQTT and AMQT protocols and the various supported modules (including Machine Learning, Stream Analytics or IoT Hub) are packed and deployed as Docker containers on top of IoT Edge.

 Google is the latest provider to add IoT PaaS as part of its Cloud Platform with the recent announcement of Cloud IoT Core. The offering consists of two main frameworks:
 \begin{itemize}
     \item a device manager service responsible for initially registering each IoT instance establishing identity, along with an authentication mechanism
     \item a messaging bridge built on MQTT protocol able to collect data from customer devices and deliver them to Google's Cloud Pub/Sub service. 
 \end{itemize} 
 The Pub/Sub service messages in high volume over HTTP or gRPC \cite{gRPC}, while supported languages include Java, Go, .NET, JavaScript, C, Python, Ruby, and PHP. In addition, Cloud IoT Core is highly integrated with Cloud Functions (serverless capabilities), Cloud Dataflow (real-time or batch data processing), and Cloud Machine Learning (predictive analytics), while datasets can be stored in BigQuery.

Regarding IBM, the fully managed, cloud-hosted service that makes it simple to derive value from IoT devices is the Watson IoT Platform. It supports connections through the MQTT messaging protocol including a maximum of 500 connected devices, with data exchange and analysis limit of 200 MB for each. IBM provides a variety of solutions carefully mapped to the need of different industries including Automotive, Electronics, Banking, and Retail. Such offerings are deeply empowered by the cloud-cognitive capabilities added by the Watson platform and the vendors's data-first approach that is adaptable to businesses of all kinds. Recently, IDC MarketScape's IoT vendor 2017 assessment has singled out IBM's offering with Watson availability, instant cognitive analytics and the company's security strategy being the tip of the spear \cite{idcvendor}.

\section{Future Research Challenges and Directions}
\label{section:chal}

The massive involvement of various cloud industry players coupled with the emerging IoT and smart environment paradigms (that are creating massive amounts of data under management and processing) is bound to create new requirements and open new research directions. Cloud economics research is of great interest to cloud vendors revolving around service pricing, user activity monitoring, and vendor financial agreements described by SLAs (relate vendor supply and customer resource demands \cite{faniyi2016systematic}). A related challenge concern inter-vendor migration of existing services that affect provisioning methods, and availability guarantees. Future research focuses on reducing the deployment overhead with the use of the serverless computing paradigm and container-based approaches moving away from VMs to support intensive real-time workloads, and per function-activation charging. All discussed cloud vendors offer their flavor of this option (Amazon Lambda, Azure Functions, Google Cloud Functions, IBM Bluemix OpenWisk). Still, the big-scale deployment of the paradigm that includes appropriate scaling, managing transaction rates, and heterogeneous hardware adaptation and benchmarking is under discussion \cite{eivy2017wary}.

Current cloud datacenter geographical distribution over distant regions creates data replication challenges despite the latest attempts to mitigate issues related to latency, consistent replicas, and retaining low response times. Migration between different types of databases and hard code schema definitions can also become a problem. Many cloud users and vendors have been focusing on how to achieve high SLA even under failures. Failures on the Cloud are becoming common, hence performing Chaos exercises is becoming the new norm \cite{basiri2016chaos}. Chaos Engineering is the discipline of experimenting on a distributed system to build confidence in the system’s capability
to withstand turbulent conditions in production. Open challenges include the integration of appropriate caching approaches and architectures (AWS's DAX is a proposal towards this direction), and database services' benchmarking during workload peaks. Also, the related reliability threats (temporal/spatial correlated failures) due to the interconnectivity and scale of modern data centers can be mitigated with fault tolerance improvements on cloud storage to handle Big Data applications \cite{nachiappan2017cloud}. Apart from that, failure characterization and prediction models based on deep learning are emerging to provide guarantees regarding performance reliability and are extended to account for Fog computing deployments and IoT-related edge components.

Moreover, a major research challenge is cloud interconnection and interoperation. In the industry many small to mid-size entities have focused on systems that can assist them to provide public and on-premise products like Kubernetes for automating deployment, scaling, and management of containerized applications, or Spinnaker as a multi-cloud continuous delivery platform \cite{spinnaker}. Research has also focused on combining functionalities from different providers towards fulfilling customer cost/resource constraints via composition of services. Another research challenge revolves around cloud infrastructure scalability needs, that is practically limited by the scalability of individual components including storage, computing nodes, and networking. Research is focusing on dedicated cloud deployments that examine specialized applications (e.g., machine learning, image recognition) while cloud vendors already offer specialized hardware for these purposes such as Google's Tensor Processing Units \cite{jouppi2017datacenter}, Amazon's GPUs offering, and Microsoft's inclusion of FPGAs in Azure cloud \cite{putnam2014reconfigurable}. A comprehensive study of challenges and future directions that will concern cloud computing research during the next decade can be found in \cite{buyya2017manifesto}.

The major cloud service providers (Amazon, Google, Microsoft, and IBM) will continue to innovate with new cloud-based services. These additions are, in essence, heavily influenced by the demands, and directions of other industries that rapidly invest in cloud-based solutions, e.g., healthcare \cite{kaur2014cloud} and automobile \cite{he2014developing}. These services are more likely to be centered around IoT, microservice architectures, containers, cross-cloud data management, cloud-based artificial intelligence integration between machine and humans (e.g., Microsoft's Cortana, Amazon's Alexa, Google's Cloud Machine Learning Engine, Apple's Siri). These services will need large-scale computing, storage, and functionality in new form factors that will integrate with our everyday life (e.g., wearables, vehicles). Because of the new emerging applications that have low-latency and high-bandwidth requirements, cloud vendors will continue to invest in deploying datacenters across different places worldwide.

These examples showcase how the major cloud vendors are aggressively expanding the market towards different emerging areas. Hence, one can only capture a point in time. The collected and presented information pertains to a specific time frame including updates announced up to late 2017.

\section{Conclusion}
\label{section:con}

In this paper, we conduct a taxonomy and survey of cloud services offered by four dominant, in terms of revenue, cloud infrastructure vendors. We map the cloud-based services into several major categories: computing, storage, databases, analytics, data pipelines, machine learning, and networking services. For each family, we present the services currently offered along with the associated characteristics, and the features that separate each vendor. Regarding computing, storage, and networking, all cloud vendors offer strong products in terms of functionality (provided that pricing is not a variable), as these categories are the core of cloud computing and have been thoroughly developed into mature services. On the other hand, there is a variety of different choices concerning databases, data analytics products and AI support. All four providers provide impressive no-sql, relational, and petabyte-scale data warehouse offerings and services with similar characteristics concerning data processing and orchestration, building blocks, streaming capabilities and machine learning.

\ifCLASSOPTIONcaptionsoff
  \newpage
\fi

\bibliographystyle{IEEEtran}
\bibliography{IEEEabrv,references}

\end{document}